\documentclass[fleqn,10pt]{wlscirep}
\usepackage[utf8]{inputenc}
\usepackage[T1]{fontenc}
\usepackage{graphicx}
\usepackage{dcolumn}
\usepackage{bm}
\usepackage{xcolor}
\usepackage{physics}
\usepackage{pifont}
\usepackage{verbatim}
\usepackage{fancyvrb}

\usepackage{microtype}
\title{Carrier–Envelope Phase Control of Orbital Angular Momentum in Solid-State High-Harmonic Generation}
\usepackage{xcolor}

\author[1]{Camilo Granados}
\author[2]{Rajaram Shrestha}
\author[3]{Bikash Kumar Das}
\author[2]{Debobrata Rajak}
\author[2,4]{Eric Cormier}
\author[2]{Bálint Kiss}
\author[5]{Carmelo Rosales-Guzman}
\author[1,*]{Wenlong Gao}

\affil[1]{Eastern Institute of Technology, Ningbo 315200, China}
\affil[2]{ELI ALPS, The Extreme Light Infrastructure ERIC, Wolfgang Sandner u. 3., 6728 Szeged, Hungary.}
\affil[3]{Theoretisch-Physikalisches Institut, Friedrich-Schiller-Universität Jena, Max-Wien-Platz 1, D-07743 Jena, Germany}
\affil[4]{Laboratoire Photonique Numérique et Nanosciences (LP2N), UMR 5298, CNRS-IOGS-Université Bordeaux, 33400 Talence, France}
\affil[5]{Centro de Investigaciones en Óptica, A.C., Loma del Bosque 115, Colonia Lomas del campestre, 371507 León, Gto., Mexico}
\affil[*]{wgao@eitech.edu.cn}


\begin{abstract}
High-order harmonic generation (HHG) driven by optical vortices is a powerful route to produce structured light in different spectral regions. The nonlinear process transfers orbital angular momentum (OAM) from the driving field to the emitted harmonics according to the scaling law $l_q = q\times l$, a consequence of the rotational invariance and angular momentum conservation. Here, we show that, in the regime of few-cycle pulses, the topological charge (TC) of the harmonic radiation detected within a finite spectral window is no longer fixed by this scaling law alone, but is governed by the interplay between broken crystal inversion symmetry and carrier-envelope phase (CEP)-sensitive sub-cycle electron dynamics. By driving HHG in a ZnO crystal with few-cycle ($\approx 1.5$ cycles) vortex beams centered at 3.2~$\mu$m, we observed that the measured TC becomes strongly CEP-dependent, switching between adjacent integer values, but only when the inversion symmetry is broken and the harmonic emission is CEP-sensitive. The TC switching vanishes when either condition is removed. Numerical analysis reveals that the TC switching originates from a CEP-controlled redistribution of spectral weight among spectrally overlapping harmonic orders, which changes the dominant OAM channel within the detection window. These results identify the CEP as a degree of freedom for tailoring the topological structure of high-harmonic radiation, pointing toward waveform-controlled structured attosecond light sources. 
\end{abstract}

\begin{document}

\flushbottom
\maketitle
%
%
\thispagestyle{empty}


\section*{Introduction}

Symmetry principles play a central role in modern physics, establishing conservation laws that govern the behavior of physical systems \cite{Nother}. In atomic and optical processes, rotational symmetry is directly linked to the conservation of angular momentum, providing a powerful framework for understanding the interaction of structured electromagnetic fields with matter \cite{Allen_OAM}. Investigating the response of strongly driven systems under conditions where these symmetries are modified or dynamically controlled offers unique opportunities to explore new regimes of light-matter interaction \cite{Cohen_Symmetries, Helical_dichro}.

Light beams possessing OAM \cite{Allen_OAM, HHG_OAM_EXP, Corkun1}, a prominent example of structured light known as optical vortices, are typically equipped with a helical phase of the form $\exp(i l \phi)$ in their field distribution, where $l$ is an integer known as the TC and $\phi$ is the azimuthal coordinate. Each photon in such a beam carries an OAM of $l\hbar$. In quantum-mechanical terms, the uncertainty relation between angle and angular momentum dictates that a well-defined value of the OAM can only be achieved when there is no angular restriction i.e., the beam's intensity distribution must exhibit circular symmetry. Additionally, the rotational symmetry of the optical beam ensures that well-defined angular momentum conservation laws apply during nonlinear optical processes \cite{HHG_OAM_EXP,Corkun1}.

High-order harmonic generation (HHG) driven by structured light has emerged as a versatile platform for extending these conservation laws into new nonlinear spectral regimes \cite{HHG_OAM_EXP2,HHG_OAM_EXP5}. Experiments in both atomic gases \cite{HHG_OAM_EXP, Corkun1} and solids \cite{OAM_solids, SAM_OAM_GaSe, SHG_OAM} have demonstrated that the OAM carried by the driving field is transferred to the emitted harmonics according to the scaling law $l_q = q\times l$ \cite{Corkun1, Garcia_OAM}, where $l_q$ and $l$ denote the OAM of the $q^\text{th}$ harmonic order and the driving field, respectively. This linear upscaling law for OAM applies specifically to HHG driven by a single laser field. Furthermore, it has been observed over a broad spectrum of experimental conditions and has also been confirmed theoretically. This establishes OAM conservation as one of the fundamental properties of harmonic generation driven by vortex beams \cite{Review_Bikash}. 

At the same time, the generation of high harmonics using few-cycle laser pulses has revealed the crucial role played by CEP \cite{CEP_control3, single-cycle_2022, RajaSC}. In this regime, the CEP determines the sub-cycle waveform of the driving electric field and provides direct control over the microscopic electron dynamics responsible for harmonic emission and attosecond pulse generation \cite{Attoclock, CEP_control3}. CEP-driven control of harmonic intensity, spectral structure, polarization, and emission timing has been extensively investigated in both atomic and solid-state systems \cite{Subcycle1, Subcycle2, Subcycle3, Subcycle4, Subcycle5, CEP_Monitoring}. Recent theoretical studies have further suggested that few cycle OAM pulses-driven harmonic generation can indeed manipulate the OAM content of the generated harmonic vortices in gases \cite{CEP_Effect_gases, ghimire_tutorial_2026}. However, there is no experimental evidence of such a novel effect either in gases or solid-state targets. 

The robustness of the linear OAM upscaling law raises a fundamental question. While rotational symmetry and angular momentum conservation fix the OAM of each harmonic order, no temporal degree of freedom has yet been demonstrated that can alter this inherent constraint. Yet few-cycle HHG is governed by sub-cycle electron dynamics that the CEP controls directly \cite{CEP_control1,CEP_control2}. Can a purely temporal degree of freedom then control the topological structure of the detected harmonic field? 

Here, we address this question experimentally by driving the harmonic generation process in ZnO with few-cycle vortex beams. We find that control over the measured TC of the detected harmonic radiation emerges only when broken crystal inversion symmetry is combined with CEP-sensitive sub-cycle dynamics. The breaking of spatial inversion symmetry enables the generation of even harmonics, reducing the spectral separation between adjacent harmonic orders and allowing mode overlap. Microscopically, CEP variations modulate the emitted harmonic radiation by controlling the sub-cycle carrier dynamics and, specifically in solids, the time at which carriers are generated. Consequently, in the few-cycle regime, the measured TC becomes strongly dependent on the CEP, switching between adjacent values. This effect is observed for an a-cut ZnO (aZnO) crystal, however, it vanishes when harmonic generation is driven in a c-cut ZnO (cZnO) crystal or with longer driving pulses, for which the emission becomes effectively CEP-insensitive. We identify that, macroscopically, the TC switching originates from a CEP-controlled redistribution of spectral weight among spectrally overlapping harmonic orders, which modifies the dominant OAM contribution of the detected harmonic field. The degree of overlap is governed by the breaking of spatial inversion symmetry in the ZnO crystal. These results establish CEP as an additional degree of freedom for controlling the topological structure of high-harmonic radiation, opening a route toward waveform-controlled structured light sources.

\section*{Results and discussion}
This section begins by discussing the theoretical framework for the CEP control of the harmonic radiation TC. Subsequently, we provide a concise overview of the experimental setup and present the experimental results. Specifically, we will show how the CEP can be used as an effective tool to control the TC and delineate the specific conditions necessary for observing this effect. 

\subsection*{Theoretical framework}

To determine how the CEP can control the harmonic radiation TC, we first calculate the harmonic field generated by a vortex driver and then evaluate the effect on a spectral filtered spectral region in the near-field. The detected radiation is obtained by extending the thin-slab model (TSM) description of vortex-driven HHG \cite{Garcia_OAM, Granados_2025} to calculate the resulting far-field topological structure. The driving field is modeled as a linearly polarized vortex pulse:
\begin{equation}
    \mathbf{E}(\mathbf{r}_\perp,t)=A(t)|R_l(\rho)|e^{i l \phi}
    \cos(\omega_0 t+\varphi)\hat{\mathbf{e}},
    \label{eq:driver}
\end{equation}
where $A(t)$ is the temporal envelope, $\varphi$ is the CEP, and $\omega_0$ is the carrier angular frequency. The transverse amplitude of the beam is given by $R_l(\rho)$, while the term $\exp(il\phi)$ accounts for the helical phase of the vortex beam, where $l$ is the TC. Furthermore, $\mathbf{r}_\perp=(\rho,\phi)$ are the radial and azimuthal coordinates of the beam, respectively, and $\hat{\mathbf{e}}$ denotes the polarization vector. Since electron dynamics are confined to the polarization direction of the linearly polarized laser field, we drop the polarization vector in what follows.

The helical phase of each harmonic order is fixed by the rotational symmetry of the interaction: the generation of the $q^\text{th}$ harmonic involves the absorption of $q$ photons from the driving field, each carrying an OAM of $l\hbar$, and therefore imprints a phase factor (helical phase) $e^{i q l \phi}$ to the individual harmonic field. Within the TSM description of vortex beam-solid nonlinear interactions \cite{Granados_2025}, the near-field phase of the $q^\text{th}$ harmonic is expressed as the sum of two distinct contributions: (1) $q$ times the phase of the fundamental driving field, and (2) an intrinsic, material-dependent phase governed by the nonlinear response coefficients and the driving field intensity. Similarly, the harmonic amplitude scales with the fundamental driving field amplitude raised to the power $p$, where $p$ represents the nonlinear scaling factor of the process and is conventionally extracted from the harmonic intensity scaling law. Consequently, the complex harmonic amplitude in the near-field can be written as:
\begin{equation}
    E_q(\mathbf{r}_\perp,t)\propto A_q(t)\big|R_{ql}(\rho)\big|^p e^{i q l \phi}
    e^{i\left(q\omega_0 t + q\varphi + \phi_\alpha \right)},
    \label{eq:harmonic}
\end{equation}
with $\phi_\alpha$ representing the intrinsic CEP-independent phase. The phase terms $\exp(i q l \phi)$ and $\exp(i q\omega_0 t)$ in Eq.~\eqref{eq:harmonic} ensure the conservation of OAM ($l_q=q\times l$) and energy ($\omega_q=q\times\omega_0$) in the harmonic generation process, respectively, and are a direct consequence of the dipole approximation. Additionally, the CEP contribution of the $q^\text{th}$ harmonic is given by $q\times\varphi$. Therefore, the OAM carried by each individual harmonic order satisfies $l_q=q\times l$ and remains independent of the CEP.

For few-cycle driving pulses, however, the CEP modifies the sub-cycle electron dynamics. In particular, the asymmetric contribution of electron trajectories launched during different half cycles changes the amplitude and phase of the emitted harmonic components \cite{Subcycle4}. To calculate the harmonic radiation, we consider the coherent superposition of two spectrally consecutive harmonic orders in the far-field (see the Methods section):
\begin{eqnarray}
    E_w(\mathbf{r}_\perp,\omega;\varphi)&\propto&
    a_q(\varphi)|R_{ql}(\rho)|^{p}e^{i q l \phi}
    +a_{q+n}(\varphi)|R_{(q+n)l}(\rho)|^{p}e^{i(q+n)l\phi}
    e^{i\Delta\phi_{\rm rel}(\varphi)}, \nonumber\\
    &\propto&
    a_q(\varphi)\Bigg(
    |R_{ql}(\rho)|^{p}e^{i q l\phi}
    +
    \frac{a_{q+n}(\varphi)}{a_q(\varphi)}
    |R_{(q+n)l}(\rho)|^{p}
    e^{i(q+n)l\phi}
    e^{i\Delta\phi_{\rm rel}(\varphi)}
    \Bigg).
    \label{eq:window}
\end{eqnarray}

Here, $a_q(\varphi)$ denotes the detected far-field harmonic amplitude withing a the finite spectra region where the modes overlap. Its CEP dependence, originating from sub-cycle electron dynamics, determines the relative weight of the different OAM-carrying harmonic components within the spectral window. Importantly, The relative phase between the two harmonic contributions is given by: 
\begin{equation}
    \Delta\phi_{\rm rel}(\varphi)=n\varphi+\Delta\phi^0,
\end{equation}
where $n=q'-q$, and $\Delta\phi^0$ is defined in Eq.~(\ref{eqn14}) of the Methods section. Note that the relative amplitude of the two components determines the dominant OAM contribution and consequently the measured TC.

Using the calculated far-field amplitude in Eq.~\eqref{eq:window}, the resulting TC can be obtained from the winding number analysis of the two-mode superposition field. When one of the two OAM components dominates within the detection window, the calculated TC approaches:
\begin{equation}
      \ell =
  \begin{cases}
    q\times l, & |a_q| |R_{ql}(\rho)|^{p}>|a_{q+n}| |R_{(q+n)l}(\rho)|^{p},\\[2pt]
    (q+n)\times l, & |a_{q+n}| |R_{(q+n)l}(\rho)|^{p}>|a_q| |R_{ql}(\rho)|^{p}.
  \end{cases}
  \label{eq:lw}
\end{equation}
This result predicts that the measured TC is determined by the dominant OAM contribution within the detected spectral window. Since the CEP controls the relative contribution of electron trajectories launched during different half cycles of the driving field, few-cycle pulses produce a CEP-dependent redistribution of the neighboring harmonic amplitudes. Consequently, the dominant OAM contribution within the detection window can switch with the CEP. A detailed calculation of Eqs.~(\ref{eq:window}) and (\ref{eq:lw}) is provided in the Methods section.

The effect of inversion-symmetry breaking is calculated by solving the semiconductor Bloch equations (SBEs) including a complex transition dipole moment, following the approach introduced in Refs.~\cite{EvenH_dipoleZnO, EvenH_dipoleGraphene}. Starting from the conventional one-dimensional SBEs for a two-band system, consisting of one conduction and one valence band, we introduce a complex interband dipole $d(k)=|d(k)|e^{i\Phi(k)}$, whose phase accounts for inversion-symmetry breaking \cite{EvenH_dipoleZnO, EvenH_dipoleGraphene}. This modification leads to coupled equations for the interband coherence and carrier populations, which are solved numerically to obtain the interband polarization, intraband current, and HHG spectrum.

The calculated HHG spectra reveal that CEP-dependent redistribution of neighboring harmonic orders occurs only when the complex dipole is included, whereas the conventional real-dipole model predicts CEP-insensitive harmonic emission. These calculations therefore predict that inversion-symmetry breaking is required for CEP-controlled TC switching. By generating even harmonics and reducing the separation between adjacent harmonic orders to $n=1$, inversion-symmetry breaking enables spectral overlap within the detection window.

Our theoretical calculations predict that to achieve control over the harmonic radiation TC requires two physical ingredients. First, neighboring harmonic orders must spectrally overlap within the detection window, which occurs when the harmonic linewidth becomes comparable to the order spacing $n\omega_0$. To achieve this condition, it is necessary to use short pulses (few cycles) which produce substantial spectral overlap between neighboring harmonics, in contrast to the long pulses (several cycles). Second, the harmonic amplitudes must exhibit CEP-dependent redistribution. This originates from CEP-sensitive sub-cycle electron dynamics and is only present for few-cycle driving fields. Breaking the crystal inversion symmetry, as in aZnO, generates even harmonics and reduces the spectral separation between adjacent orders to $n=1$, thereby facilitating their overlap within the detection window. For longer pulses, the harmonic amplitudes become effectively CEP-insensitive, and no switching between TC values occurs regardless of the crystal orientation (See the Supplementary Material). We note that, for the plateau region of the harmonic spectra, the spectral separation between harmonics is reduced possibly making the effect less sensitive to the spatial inversion of symmetry breaking.

\subsection*{Experimental details}

To demonstrate the effect of the CEP on the harmonic OAM, we compressed the 4-cycle output pulsed beam from the OPCPA MIR laser system (available at the MIR laboratory, ELI-ALPS, Hungary) down to sub-two-cycle duration. The beam possesses a central wavelength of 3200~nm, a repetition rate of 100 kHz, and delivers a pulse energy of 100~$\mu$J. This compression was achieved using the experimental setup presented in Fig. 1(a)\cite{RajaSC, Short_pulses2}, which was specifically optimized to pre-compensate for the material dispersion introduced downstream by the wire grids (used to control the delivered power) and a 1-mm fused silica spiral phase plate (SPP). Following the compression stage, these beams were directed through the SPP to generate vortex beams (VBs) with a TC of $l = 1$ and radial index different from zero. Crucially, by accounting for this material dispersion, the final driving vortex beam retains a 1.5-cycles pulse duration.

\begin{figure*}[ht!]
    \centering
\includegraphics[width=1\columnwidth,trim=0cm 0cm 0cm 0cm, clip]{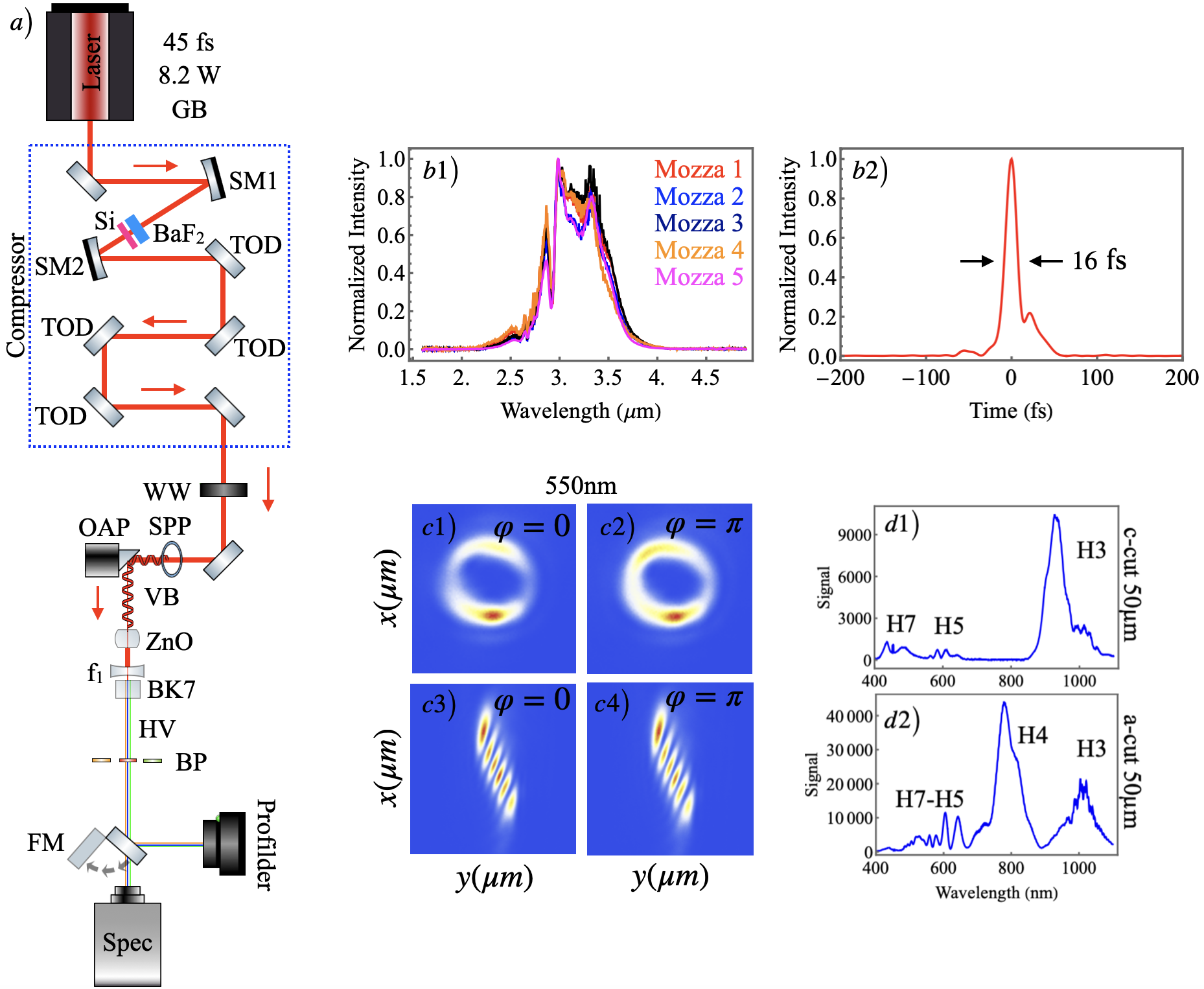}
    \caption{Experimental setup, temporal and spectral characterization of the pump, harmonic spectra obtained for ZnO with different crystal cuts, and harmonic vortices for different CEP values. The 4 cycles, 3200~nm, 40 mW and 100~kHz driving laser was compressed to achieve short pulses of $\approx 16$~fs. The short pulses were then guided towards the SPP to generate VBs. The VB was then guided towards the solid target, where an off-axis parabolic mirror (f=150~mm) focuses ($\approx$ 50~$\mu$m beam spot) the short pulses to drive the harmonic generation process. The fundamental MIR beam was blocked using a BK7 glass. The harmonic beams were guided towards the spectrometer and beam profiler for spectral and spatial characterization. In (b1), we present the pump beam spectra for different pulse durations. Mozza1 represents the pump beam after the compression stage, while Mozza2 represents the pump beam after the SPP. The chirped beams are represented by Mozza3 and Mozza4, which correspond to the pump beam after propagating through 4~mm of CaF$_2$ and 4~mm of CaF$_2$ + 2 mm of BaF$_2$, respectively. In (b2), we show the pulse duration measurement of driving vortex beam. In panels (c1)-(c4), we show the transverse intensity distributions of the 5$^\text{th}$ harmonic vortex and their corresponding tilted HG lobed-patterns generated with the long pulsed beam, and for two different CEP values. These measurements represent reference beam profiles for the case of no CEP effect. In panels (d1) and (d2), we show the harmonic spectra measured for cZnO and aZnO, when the compressed driving vortex beam is used for HHG. The CEP effect is clear in the region between the harmonics 7$^\text{th}$ and 5$^\text{th}$. In particular, the measured spectral window, given by the band-pass filter, is located around 550$\pm$~20~nm. We note that this spectral window is closer to the 6$^\text{th}$ harmonic order than the 5$^\text{th}$ order. In the figure, GB: Gaussian beam; SM: spherical mirror; Si: Silicon plate; TOD: third order dispersion mirrors; WW: wire grids; SPP: spiral phase plate; OAP: off-axis parabolic mirror; VB: vortex beam; HV: harmonic vortices; BP: band-pass filters; FM: flip mirror; Spec: spectrometer.}
    \label{Fig1}
\end{figure*}

Immediately after the SPP, several optical windows were introduced into the vortex beam path to control the temporal duration of the pump pulse by adding dispersion. The corresponding pump spectra were characterized using a Multi-Octave Spectrum Analyser (Mozza), as shown in Fig.~\ref{Fig1}(b1). The different measurements correspond to the pulse after compression (Mozza1), after the SPP (Mozza2), and after propagation through increasing amounts of dispersive material: 4~mm CaF$_2$ (Mozza3), 4~mm CaF$_2$ + 2~mm BaF$_2$ (Mozza4), and 7~mm CaF$_2$ + 6~mm BaF$_2$ (Mozza5). The spectra remain essentially unchanged for all configurations, confirming that the optical windows introduce temporal stretching while preserving the spectral bandwidth of the vortex pulse.

We measured the pulse duration of both the fundamental Gaussian and VBs after the compression stage. The optimized compressor yields short pulses of temporal duration 16~fs (FWHM), corresponding to approximately 1.5 to 1.6 optical cycles~\cite{RajaSC}. An example of the pulse-duration measurements, performed using a tunneling ionization with a perturbation for the time-domain observation of an electric field (TIPTOE) apparatus, is shown in Fig.~\ref{Fig1}~(b2). Moreover, the pulse duration after a 4mm of CaF$_2$ was calculated to be 77.2fs (7.2 cycles) and after a 7mm of CaF$_2$ + a 6mm of BaF$_2$ is 186~fs (17.4 cycles). It is important to note that generating a long pulse after producing a pulse with approximately 1.5 cycles is fundamentally different from generating the long pulses corresponding to the 4-cycle output of the MIR laser. Throughout the manuscript, we refer to these four distinct configurations as the fully compressed (16~fs), semi-stretched (77.2~fs), stretched (186 fs), and uncompressed (45~fs) driving regimes.

Following the vortex beam generation, the LG beam was directed towards the laser-solid interaction region. There, the beam was focused onto the target using an unprotected gold-coated, 150 mm focal length off-axis parabolic mirror (OAP), to induce the material's nonlinear response. The target materials investigated were aZnO (with thicknesses of 50~$\mu$m and 200~$\mu$m) and cZnO (with a thickness of 50~$\mu$m). We choose these specific materials to investigate the sensitivity of the CEP to crystal symmetry, as the broken in-plane symmetry of the aZnO strongly facilitates the generation of both even and odd-order harmonics, whereas the symmetry rules for the cZnO, under normal incidence condition, restrict the spectrum predominantly to odd-order harmonics only.

Following harmonic vortices generation in ZnO, the harmonics were isolated from the driving vortex beam using a 1-cm-thick borosilicate crown glass (BK7) element. Subsequently, the harmonic vortices were then detected with the help of an Avantes spectrometer (with a spectral range between 200 to 1260~nm) to record CEP scans. The CEP was actively set and stabilized within the optical parametric chirped-pulse amplification (OPCPA) front-end using an acousto-optic programmable dispersive filter (AOPDF; Dazzler, Fastlite). CEP stabilization was implemented through a closed-loop feedback system governed by a commercial f-to-2f interferometer (Fringeezz, Fastlite). This diagnostic tool monitored the CEP on a shot-to-shot basis at a 10 kHz repetition rate, enabling the laser system to maintain an exceptional residual CEP stability of $< 150$~mrad RMS. For the experimental scans, the CEP was systematically varied over a full 2$\pi$ cycle, spanning from -$\pi$ to $\pi$ across 31 equidistant phase steps (with an increment of 0.2 rad). 

Additionally, the transverse intensity profiles of the generated harmonic vortices, along with their corresponding TCs, were measured using a Basler camera (acA1440, spectral range 400-1000~nm). The OAMs of the harmonic vortices were characterized using a mode converter \cite{Beijersbergen1993, GranadosOAM}. The mode converter setup comprises of a single cylindrical lens, which typically produces tilted Hermite-Gaussian (HG) lobes, when circularly symmetric vortex beams illuminate it. The TC of individual harmonics or in general any emitted radiation, can be extracted from the tilted HG-lobed profiles by simply counting the number of minima between the maximum intensity lobes. Deviation from the HG pattern or un-resolved minima indicate the presence of other OAM modes. Individual harmonics or spectral windows in between the harmonics were selected using a 20-nm-bandwidth band-pass filter. Examples of the transverse intensity distributions of harmonic vortices and their corresponding tilted HG-lobed profiles are presented in Fig.~\ref{Fig1}~(c1)-(c4) for the 4 cycles long pulse, and for two different CEP values i.e., $\varphi=0$ and $\pi$. For both CEP values, the intensity distributions exhibit continuous intensity rings and are similar to each other, indicating that the CEP has no significant influence on the modification of harmonic OAM in the case of long driving pulses. In addition, examples of the harmonic spectra obtained from aZnO and cZnO are shown in Fig.~\ref{Fig1}~(d1) and (d2), respectively. Due to symmetry reasons, the generation of the $4^{th}$ harmonic is suppressed in the cZnO, whereas the same harmonic is clearly visible in the spectrum obtained from aZnO. In panel (d2) the action of the CEP is evident by the generation of multiple peaks in the spectral region between harmonics $5^{th}$ and $7^{th}$. Notice that both spectra were generated with a short pulse vortex driving field (1.5 cycles). 

\subsection*{CEP effect on the topological charge}
\begin{figure}[ht!]
    \centering
    \includegraphics[width=1\columnwidth,trim=0cm 0cm 0cm 0cm, clip]{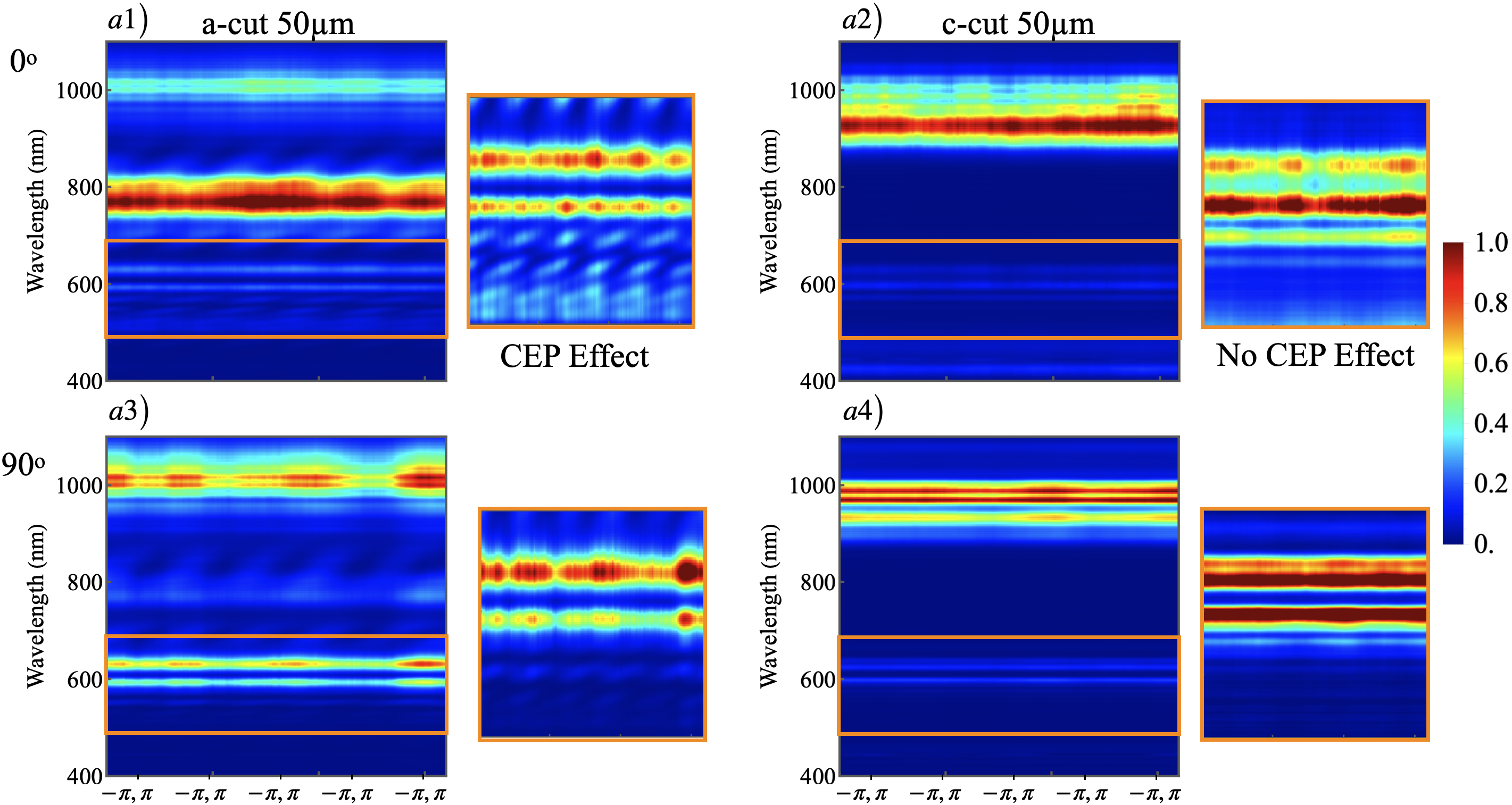}
    \caption{Carrier-envelope phase measurements. In (a1) and (a3), we show the measured CEP scans for aZnO corresponding to two different crystal orientations. The angles ($0^{\circ}$ and $90^{\circ}$) were selected for breaking and preserving the spatial inversion symmetry, respectively. The scans reveal CEP-dependent spectral features for the regions between the main harmonic peaks, which corresponds to H3 ($\approx$1066~nm), H4 ($\approx$800~nm), H5 ($\approx$640~nm), and H6 ($\approx$533~nm). The corresponding results for cZnO are presented in (a2) and (a4), for identical crystal orientations. In contrast to the case of aZnO, CEP effect is not observed in cZnO. For all the measurements presented here, we used the fully compressed driving pulse (16~fs). The $x$-axis in all the figures represents the CEP values. }
    \label{Fig2}
\end{figure}

To observe the CEP effect on the harmonic radiation, we recorded, for each CEP value, a harmonic spectra. An example of the measured CEP scans is presented in Fig.~\ref{Fig2}, for both aZnO and cZnO. We considered two distinct crystal orientation angles i.e., $0^{\circ}$ and $90^{\circ}$, for each of the two crystals, relative to the driving field's polarization direction. In all the panels, we present 5 consecutive CEP scans (from $-\pi$ to $\pi$ ). The $-\pi,\pi$ convention means that the scan finishes in that point with a CEP phase of $-\pi$ and starts a new scan with a CEP phase of $\pi$. The results obtained for the aZnO are presented in Figs.~\ref{Fig2}~(a1) and (a3). The results clearly reveal a CEP-dependent behavior, which is more pronounced particularly between consecutive harmonic orders. The CEP effect is particularly strong in the high harmonics region, as demonstrated by the insets besides each panel. The presence of diagonal maxima constitutes the characteristic fingerprint of the CEP effect, thereby revealing that the CEP modulates the superposition of harmonic vortices at the measured wavelengths. In contrast, for the case of cZnO, the CEP effect is absent from the CEP scans, as shown in Figs.~\ref{Fig2}~(a2) and (a4). These first set of measurements demonstrate not only the CEP sensitivity of the harmonic radiation when the driving field is composed of few cycles, but also the necessity of having even-order harmonics to induce the CEP effect with the 1.5 cycles driving vortex beam. 

\begin{figure}
    \centering
    \includegraphics[width=0.6\columnwidth,trim=0cm 0cm 0cm 0cm, clip]{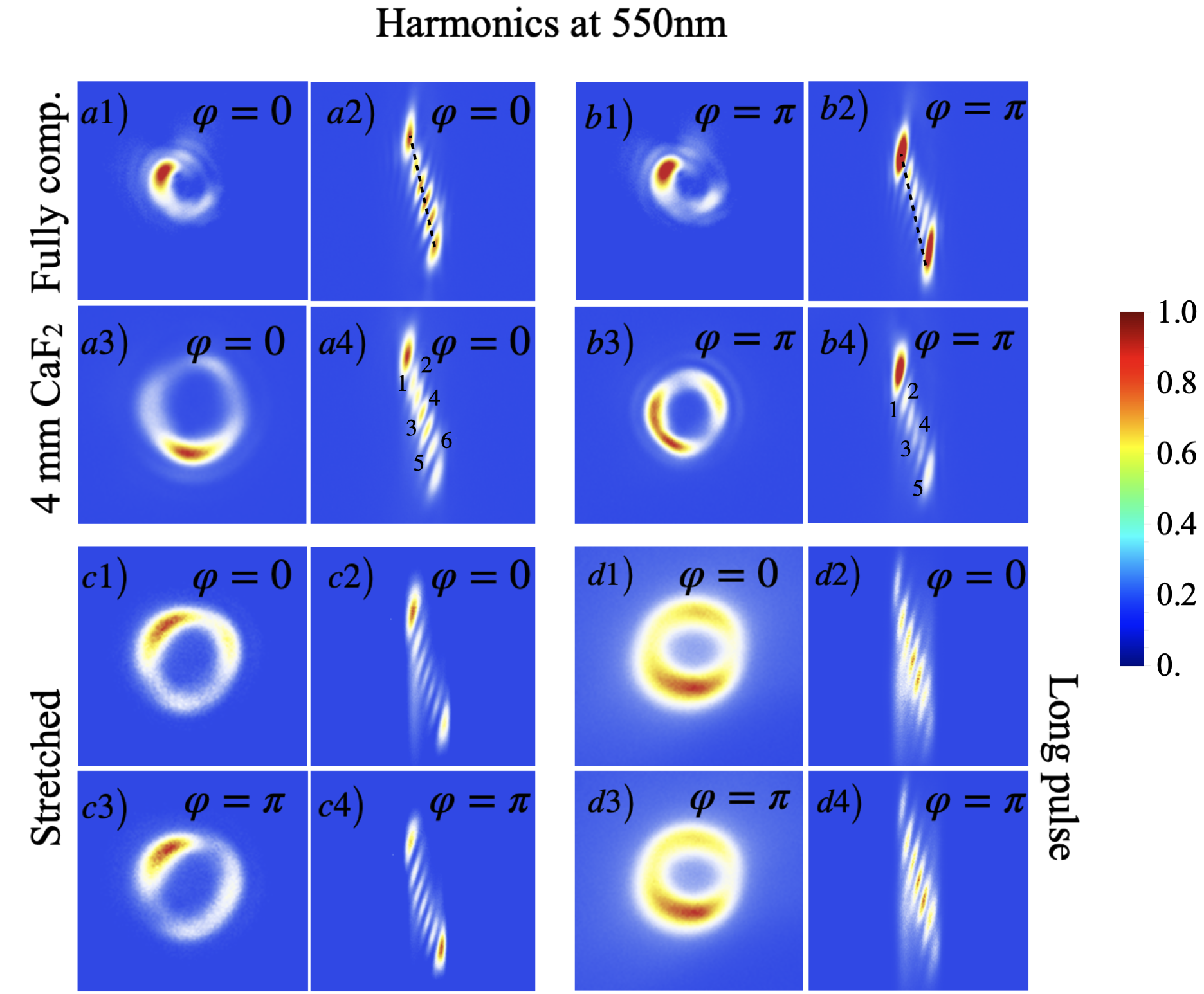}
    \caption{Harmonic vortices and TC measurements for a 200~$\mu$m-thick aZnO. In (a1) and (b1), we present the measured harmonic vortex beams generated with a fully compressed beam and for $\varphi = 0$ and $\pi$, respectively. The corresponding TC measurements are shown in (a2) and (b2). In the panels, the dashed lines indicates the effect of variying the CEP, which changes not only the lobes distribution but also the tilt of the pattern. In (a3) and (a4), we present the semi-stretched ( 4mm~CaF$_2$, 77.2~fs) case for $\varphi = 0$, while in (b3) and (b4) we show the case for $\varphi = \pi$. In the panels, we indicate the number minima between the lobes with corresponds to the TC value of the harmonic radiation. The measurements show a clear CEP effect detected in different TC values. Since the driving vortex beam possesses a TC of $l=1$, the harmonic vortices should have TCs of $l_q=5$ and $l_q=6$ for the 5$^\text{th}$ and 6$^\text{th}$ harmonic orders, respectively. For the stretched driving field (186~fs), presented in (c1) to (c4), the multiple mode character of the vortex was not observed in either the harmonic vortices (with no rotational symmetry) or the TC measurements. Importantly, for this beam configuration, the CEP effect is highly reduced. In (d1) to (d4), we present the case for a 4 cycles driving pulse (long pulses), where we observed no CEP effect.} 
    \label{Fig3}
\end{figure}

We now turn our attention to the harmonic vortices beam profile and their topological charge measured through HG intensity patterns, produced with a cylindrical lens, and for different pulse lengths. These measurements demonstrate the influence of the CEP on both the intensity distribution and the TC content of the generated harmonic vortices, as we will show below. Examples of the measurements are shown in Fig.~\ref{Fig3}. Panels (a1) and (a2) display the beam profile and TC measurements, respectively, recorded at a wavelength of 550~nm for $\varphi=0$ under a fully compressed driving field. This wavelength falls between the fifth and sixth harmonic orders. The corresponding results for $\varphi=\pi$ are shown in panels (b1) and (b2). As evident from Figs.~\ref{Fig3}~(a1), (a2), (b1), and (b2), both the harmonic vortex intensity distribution and the HG lobed pattern undergo noticeable changes with variations in the CEP of the driving field. The changes are indicated by the dashed black lines, which show a change in the tilt of the HG pattern accompany by changes in the intensity distribution of the HG lobes. Interestingly, the spatial profile of the harmonic vortices indicates the emergence of fractional harmonic vortices, as the intensity distribution no longer exhibits rotational symmetry. This observation is corroborated by the TC measurements, which reveal multiple lobe structures surrounding the dominant HG lobes, along with noticeable variations in the intensity distribution (see the Supplementary material). We note that, in the visible regime and using continuous-wave laser sources, it has already been demonstrated that fractional vortices typically exhibit a dark radial opening in their intensity ring. This indicates a breaking of the radial (or circular) symmetry of the vortex structure. The generation of fractional vortices through solid-state HHG represents an important advancement, as it enables access to extremely short-wavelength fractional vortices through the generation of very high-order harmonics. This, in turn, can significantly extend the applicability of fractional vortices into the few-nanometer wavelength regime.

To further investigate the CEP dependence and its relationship with the temporal-spectral distribution of the pulse, the fully compressed beam was stretched using a 4 mm-thick CaF$_2$ window. We present in Figs.~\ref{Fig3} (a3) and (a4), the results for $\varphi=0$, while the corresponding results for $\varphi=\pi$ are shown in Figs.~\ref{Fig3} (b3) and (b4). The TC measurements reveal a clear CEP-induced modification in the TC, changing from $l_q\approx6$ for $\varphi=0$ to $l_q=5$ for $\varphi=\pi$. This observation clearly demonstrates the sensitivity of the harmonic-vortex OAM content to the CEP. Note that since the TC of the driving field is $l=1$, an integer harmonic should follow the scaling law $l_q=q\times l$. Accordingly, the 5$^\text{th}$ and 6$^\text{th}$ harmonic vortices should possess TC values of $l_q=6$ and $l_q=5$, respectively. Also, note that the resulting harmonic vortex beam profiles are in good agreement with the theoretical calculations for a coherent superposition of different modes (see Supplementary Material). 

The observation of CEP-dependent OAM distributions raises an important question: if the sub-cycle strong-field dynamics redistribute the OAM content across multiple harmonic modes, is it still possible to generate harmonic vortices with a well-defined single TC in the ultrashort-pulse regime? To address this question, we employed a stretched driving vortex beam (17.4 cycles) to generate harmonics at the same wavelength. The resulting beam profiles, together with their corresponding TC measurements are presented in Fig.~\ref{Fig3} (c1)-(c4). Only minor variations are observed in both the vortex intensity distributions and the corresponding HG profiles. Moreover, no change is observed in the measured TC. These observations indicate that the pulse duration becomes sufficiently long to effectively suppress the CEP-sensitive sub-cycle dynamics. Additionally, unlike the case of a fully compressed driving vortex field, we observe a near-perfect continuous ring for the harmonic vortex in this case.

We performed the same measurements using a 4-cycle long driving vortex beam with not dispersion material in the driving field path. The results are shown in Fig.~\ref{Fig3}(d1)-(d4). These results demonstrate that harmonic vortices carrying a well-defined single TC can be generated when the harmonic spectra exhibit only limited overlap between neighboring harmonic orders. Clearly, the result is limited to the spectral region investigated here and the sensitivity of our detection system. This is an important distinction when compared with the long pulses output from the MIR laser, since this particular pulse does not contain the same spectra as the long pulses generated with the different material windows. 

Our experimental results demonstrate that the CEP modulates the harmonic emission when the driving pulse is in the few-cycle regime, as already seen from the CEP scans in Fig.~\ref{Fig2}. This is clear from the rotationally asymmetric profiles obtained with the fully compressed pulse, as shown in Figs.~\ref{Fig3} (a) and (b). For the compressed pulse, however, the spectrum is sufficiently broad that several harmonic orders overlap within the detection window. As a result, the observed vortex exhibits a fractional character, preventing the unambiguous assignment of a single integer TC. Introducing a moderate chirp narrows the overlap region such that only two adjacent harmonics superimpose within the band-pass window. The same CEP-driven redistribution consequently manifests as a clean switch of the measured TC between adjacent integers, as presented in Figs.~\ref{Fig3} (a4) and (b4). Further stretching of the driving pulse suppresses the CEP sensitivity entirely (see Fig.~\ref{Fig3} (c)), while the 4-cycle output lacks the spectral overlap required for any switching (see Fig.~\ref{Fig3} (d)). Note that the chirp does not alter the pump pulse power spectrum (Fig.~\ref{Fig1}); instead, it broadens the pulse duration in the time domain. Due to the time-bandwidth limit, longer harmonic emission results in narrower harmonic linewidth, $\Gamma_q$. This narrowing reduces the overlap between adjacent harmonic orders within the fixed band-pass window, thereby enabling the isolation of the two selected harmonic orders.

Collectively, these measurements provide direct evidence of CEP control over the OAM content of harmonic vortices, opening the door to explore OAM sensitive electron dynamics together via changes in the CEP.

\subsection*{Crystal thickness effect}
\begin{figure}
    \centering
\includegraphics[width=.6\columnwidth,trim=0cm 0cm 0cm 0cm, clip]{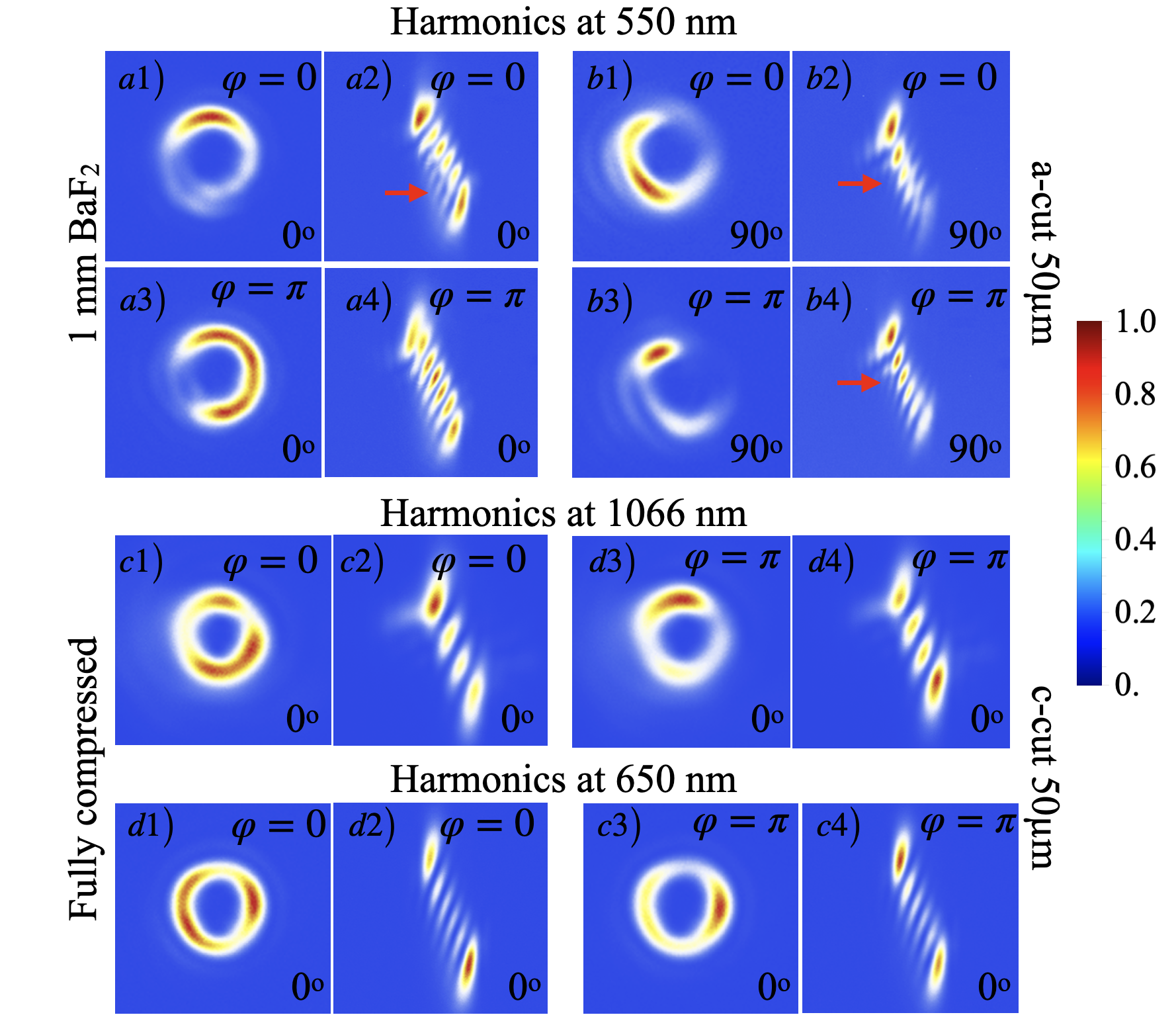}
    \caption{Harmonic vortices and TC measurements for 50~$\mu$m thick aZnO and cZnO. In (a1) to (a4), we show the vortex beam and TC measurements for two different CEP values $\varphi = 0$ and $\pi$, and for aZnO crystal oriented at an angle of $0^{\circ}$. The spectral measurement window is located around 550~nm, which is given by the band-pass filter. Interestingly, the sensitivity of the TC to the CEP is evident in the tilted HG lobed structures shown in (a2) and (a4). A similar multi-mode nature for the 550~nm window is observed for aZnO rotated by an angle of $90^{\circ}$, as shown in (b1) to (b4). Interestingly, the switching is, however, suppressed. The red lines in the panels are used to highlight the discontinuity created in the tilted HG lobes, resulting from the sensitivity to the CEP. The results for cZnO are shown in (c1) to (c4), for the third harmonic vortex, while in (d1) to (d4), for the fifth harmonic vortex. In this case, the vortex beams and the TC measurements reveal no CEP effect even when the harmonic process is driven by a fully compressed laser pulse. This supports our combined effect explanation for the CEP effect: cZnO produces odd-order harmonics only which are separated by $2\omega$, and consequently, there are no adjacent even harmonic orders to contribute to the TC switching.}
    \label{Fig4}
\end{figure} 
We further investigate on the CEP control of the detected harmonics TC by driving the HHG process in thinner aZnO and cZnO crystals. The results are presented in Fig.~\ref{Fig4}. Compared with the 200-$\mu$m-thick crystal, the 50-$\mu$m-thick crystal requires a different amount of chirp to be introduced into the driving pulse to produce the same effect. Here, the introduction of a 1-mm BaF$_2$ ($\approx 68$~fs) window reveals a pronounced CEP dependence of the TC carried by the generated harmonic vortices. In panels (a1) and (a3), it can be observed that the harmonic vortices exhibit dislocations, primarily arising from the presence of multiple OAM components. The corresponding TC measurements, shown in panels (a2) and (a4), reveal that the dominant TC changes from $l_{q}=5$ to $l_{q}\approx6$ for a CEP shift of $\Delta\varphi=\pi$. Rotating the aZnO crystal by $90^{\circ}$ suppresses the CEP-induced TC switching, as shown in panel (b1)-(b4), consistent with the symmetry breaking effect. Here, we use the following convention: For $0^{\circ}$ orientation of the crystal with respect to the laser polarization, there is generation of both even and odd harmonic orders, whereas for a $90^{\circ}$ orientation, the harmonic generation is restricted to odd orders only. Nevertheless, the presence of multiple OAM modes remains clearly evident in both the tilted HG and harmonic vortex beam profiles. These measurements consistently demonstrate a fractional OAM character for the harmonic vortices, which remains unaffected by variations in the CEP value or the crystal orientation relative to the laser polarization axis. The red arrows in different panels highlight the observed dislocations and identify the different modes present in the measurements. 

These results demonstrate that the CEP-dependence TC of the harmonic vortices is also sensitive to the crystal symmetry. When the optical axis of aZnO is aligned with the driving-field polarization, the broken inversion symmetry enables the generation of even-order harmonics, which decreases the spectral separation between adjacent harmonic orders and enhances the multi-mode nature of the detected harmonic radiation. Consequently, the measured OAM exhibits a pronounced sensitivity to the CEP. Upon rotating the crystal by $90^{\circ}$, the even-order harmonics are suppressed, leading to a significant reduction in the CEP sensitivity of the measured TC.

In contrast, when HHG is driven in the cZnO crystal, the generated harmonic vortices exhibit little or no CEP dependence for a fully compressed driving vortex beam, as shown in Fig.~\ref{Fig4}(c1)-(c4) for the third harmonic and Fig.~\ref{Fig4}(d1)-(d4) for the fifth harmonic. Furthermore, because cZnO retains inversion symmetry regardless of the crystal's orientation relative to the polarization axis of the driving field, the measurements obtained at $\varphi=0$ and $\varphi=\pi$ are nearly identical. These results further highlight the combined role of crystal inversion symmetry and CEP-sensitive few-cycle electron dynamics in controlling the measured OAM of the detected harmonic radiation. 

\section*{Conclusions} 
We showed that the high-harmonic radiation TC is strongly CEP-dependent in the few-cycle pulse regime. The effect originates from a CEP-controlled redistribution of spectral weight among spectrally overlapping harmonic orders carrying different OAM, which switches the dominant OAM contribution within the detection window. Each harmonic order individually continues to satisfy the OAM upscaling law $l_q=q\times l$. Two ingredients are required to observe this CEP control: (i) breaking the crystal inversion symmetry, which generates even-order harmonics and reduces the spectral separation between the adjacent harmonic orders to $n=1$, thereby enabling their spectral overlap within the detection window. (ii) CEP-sensitive sub-cycle electron dynamics, present only for few-cycle driving pulses, which makes the relative amplitudes of the overlapping harmonic orders CEP-dependent. Removing either ingredient, for example, by using cZnO or by stretching the driving pulse until the harmonic emission becomes effectively CEP-insensitive, suppresses the TC switching. The chirp of the driving field determines the harmonic spectral widths and, consequently, the degree of spectral overlap between neighboring harmonic orders. Furthermore, different chirp conditions are required to observe the same CEP effect in crystals of different thicknesses.

These results establish the CEP as a degree of freedom for controlling the topological structure of high-harmonic radiation. Extending the same mechanism to high-harmonic orders, where neighboring harmonics likewise spectrally overlap, would enable waveform control in the extreme-ultraviolet regime and point toward waveform-engineered structured attosecond light sources.

\section*{Methods}

In this section, we will detail the calculations supporting the experimental results presented in the main manuscript. Importantly, we based our calculations on the separability of the temporal and spatial components of the vortex beams. This assumption is supported by the dipole approximation, which is still valid in this analysis and the fact that we did not induce spatiotemporal couplings through tight focusing or any other coupling-inducing means. 

\subsection*{Band-pass filter action} 
We start by writing the harmonic field with spatial and temporal dependence as follows:
\begin{equation}
E_q(\mathbf{r}_\perp, t;\varphi)=A_q(t)\big|R_{ql}(\rho)\big|^p e^{iql\phi} e^{i(q\omega_0 t+q\varphi+\phi_\alpha)} \label{eqn5}
\end{equation}
here, $A_q(t)$ is the temporal envelope, $R_{ql}(\rho)$ is the radial amplitude of the harmonic vortex beam which depends on the TC of the driving field through $ql$ (for an VB with a zero radial index), and  $\exp(iql\phi)$ represents the helical phase term for the harmonic vortex beam. The parameter $p$ (also called the order of nonlinearity), represents the scaling of the harmonic amplitude relative to the driving field's amplitude. For a harmonic located in the perturbative (non-perturbative) regime of an HHG spectrum, $p\approx q$ ($p<q$). For each harmonic, the CEP component scales with the harmonic order as $\exp(iq\varphi)$. The separability of the spatial and temporal parts is guaranteed by the dipole approximation and neglecting spatiotemporal couplings, the harmonic field can be written in separable form. The total detected radiation is the sum over the harmonic orders whose components lie within the band-pass filter allowed bandwidth. It can be expressed mathematically by the sum over all the harmonic orders inside the particular bandwidth: 
\begin{equation}
 E(\mathbf{r}_\perp, t;\varphi)=\sum_q E_q(\mathbf{r}_\perp, t;\varphi).
  \label{eqn6}
\end{equation}

Now, since we detected the harmonic radiation in the angular frequency domain, it is necessary to calculate the Fourier transform of $  E(\mathbf{r}_\perp, t;\varphi)$. The spatial factors and constant phases can be removed from the integral, leaving only the envelope to transform as follows:
\begin{eqnarray}
  \tilde{E}_q(\mathbf{r}_\perp,\omega)
  =\big|R_{ql}(\rho)\big|^pe^{iql\phi}\,e^{i(q\varphi+\phi_\alpha)}
    \int dtA_q(t)e^{iq\omega_0 t}\,e^{-iq\omega t} =\big|R_{ql}(\rho) \big|^pe^{iql\phi}\,e^{i(q\varphi+\phi_\alpha)}
    \tilde{A}_q(q\omega-q\omega_0), \label{eqn7}
\end{eqnarray}
where $\tilde{A}_q(\Omega)=\int dt\,A_q(t)\,e^{-i\Omega t}$ is the lineshape centered at zero. Also, for simplicity we wrote $E_q(\mathbf{r_\perp},\omega) = E(\mathbf{r}_\perp,\omega;\varphi)$. We can now introduce the band-pass filter window, $W_{BP}(\omega)$, and restrict the sum in Eq.~(\ref{eqn6}) to two adjacent harmonic orders:
\begin{equation}
  \tilde{E}_w(\mathbf{r_\perp},\omega)\approx W_{BP}(\omega)\Big[
  \big|R_{ql}(\rho)\big|^p e^{iql\phi}e^{i(q\varphi+\phi_\alpha)}\tilde{A}_q(q\omega-q\omega_0) + \big|R_{q'l}(\rho)  \big|^pe^{iq'l\phi}e^{i(q'\varphi+\phi_\alpha)}
   \tilde{A}_{q'}(q\omega-q'\omega_0)\Big]. \label{eqn8}
\end{equation}
Here, $\tilde{E}_w(r_\perp,\omega)$ is the detected harmonic vortex radiation after the band-pass filter. Importantly, $W_{BP}(\omega)$ is nonzero only near $\omega_w$, so the term $W_{BP}(\omega)\tilde{A}_q(q\omega-q\omega_0)$ is appreciable only if order $q$'s lineshape reaches the window ($\omega_w$) created by the band-pass filter, i.e. $|\omega_w-q\omega_0|\lesssim\Gamma_q$. Here, $\Gamma_q$ is the linewidth of the particular harmonic $q$. For $\omega_w$ placed between two adjacent harmonic orders $q$ and $q'\equiv q+n$, only these two harmonic orders give meaningful contribution while the contribution from other harmonic orders are negligible. Now, we can define the window amplitude detected during the experiment in terms of the action of the function $W_{BP}(\omega_w)$ on the emitted harmonic spectra. The band-pass filter centered at $\omega_w$ projects both neighboring orders onto a common narrow frequency band. The transmitted contributions are nearly monochromatic at $\omega_w$, carry distinct spatial OAM and their interference is therefore stationary in time. The modes therefore interfere coherently, producing a stationary transverse pattern that survives the time-integrated detection of the camera. The complex weight with which order $q$ contributes to this pattern is $a_q(\varphi)$, defined as: 
\begin{equation}
  a_q(\varphi)\equiv \frac{1}{2\pi}  e^{i(q\varphi+\phi_\alpha)}
  \int  W_{BP}(\omega_w)\tilde{A}_q(q\omega-q\omega_0) d\omega.
  \label{eqn9}
\end{equation}
Here, for simplicity we set the measurement time to zero. The complex-valued function $a_q(\varphi)$ quantifies how much of the harmonic spectra is transmitted by the band-pass filter and its phase carries the CEP term $q\varphi$. The detected transverse field is
then a two-mode field  whose amplitudes $a_q(\varphi)$ are set by the band-pass filter: 
\begin{eqnarray}
  E_w(\mathbf{r_\perp},\omega;\varphi)&=& a_q(\varphi)\big|R_{ql}(\rho)  \big|^pe^{iql\phi}+ a_{q'}(\varphi)\big|R_{q'l}(\rho)\big|^pe^{iq'l\phi}.
  \label{eqn10}
\end{eqnarray}

At the far-field, the superposition of modes takes the form: 
\begin{equation}
E_w(\beta,\vartheta, \omega;\varphi)=\int_{0}^{\infty} \rho d\rho \int_{0}^{2\pi} d\phi E_w(\mathbf{r_\perp},\omega;\varphi) \exp \Bigg[-i\frac{2 \pi}{\lambda_q} \rho \tan(\beta)\cos(\vartheta-\phi)\Bigg]. \label{eqn11}
\end{equation}
Here, $(\beta,\vartheta)$ represent the far-field coordinates. Specifically, $\beta$ represents the harmonic divergence in the far-field, while $\vartheta$ denotes the azimuthal angle. Although the angular integral in Eq.~(\ref{eqn11}) is performed over $\phi$, the far-field amplitude is still modulated by the CEP-dependent harmonic amplitude terms $a_q(\varphi)$ and $a_{q'}(\varphi)$. 

\subsection*{The CEP effect}
To understand the effect of CEP on the detected harmonic radiation, it is necessary to separate the amplitude and phase terms in Eq.~(\ref{eqn9}) explicitly: 
\begin{eqnarray}
  a_q(\varphi)&=&|a_q(\varphi)|\,e^{i(q\varphi+\chi_q)},\nonumber \\
  a_{q'}(\varphi)&=&|a_{q'}(\varphi)|\,e^{i(q'\varphi+\chi_{q'})}, \label{eqn12}
\end{eqnarray}
where we define $\chi_q\equiv\phi_\alpha+\arg \big[1/2\pi \int W_{BP}(\omega_w)\tilde{A}_q (q\omega-q\omega_{0}) d\omega\big]$. Basically, $\chi_q$ represents the combined phase contributions of the individual harmonics, excluding the CEP-dependent phase term $q\varphi$. Substituting Eq.~(\ref{eqn12}) into Eq.~\eqref{eqn10}, and factorizing the phase terms, we can write: 
\begin{eqnarray}
E_w(\mathbf{r_\perp},\omega;\varphi) &=&e^{i(q\varphi+\chi_q)}\Big[
  |a_q(\varphi)| \big|R_{ql}(\rho) \big|^pe^{iql\phi} \nonumber \\
  &+&|a_{q'}(\varphi)| \big|R_{q'l}(\rho) \big|^pe^{iq'l\phi}\,
   e^{i[(q'-q)\varphi+(\chi_{q'}-\chi_q)]}\Big]. \label{eqn13}
\end{eqnarray}
We can now use $n=q'-q$ to define:
\begin{eqnarray}
  \Delta\phi_{\rm rel}(\varphi)&=&(q'-q)\varphi+(\chi_{q'}-\chi_q)=n \varphi+\Delta\phi^{0},\nonumber \\ 
  \Delta\phi^{0}&\equiv&\chi_{q'}-\chi_q.
  \label{eqn14}
\end{eqnarray}
The pre-factor, $e^{i(q\varphi+\chi_q)}$, basically represents a global phase. It is spatially uniform, and therefore does not modify the azimuthal phase winding. Consequently, we neglect this term in the final expression:
\begin{eqnarray}
E_w(\mathbf{r_\perp},\omega;\varphi)&\propto& a_q(\varphi)\big|R_{ql}(\rho) \big|^pe^{iql\phi} \nonumber \\
  &+&a_{q+n}(\varphi) \big |R_{(q+n)l}(\rho) \big|^pe^{i(q+n)l\phi}e^{i\Delta\phi_{\rm rel}(\varphi)} \nonumber \\ \label{eqn15}
\end{eqnarray}

Here, we simplify the notation $|a_q|\to a_q$. Interestingly, the relative phase contains the CEP-dependent contribution $n\varphi$, which modifies the interference between neighboring harmonic orders. However, as shown below, changes in the measured TC are primarily associated with the CEP dependence of the amplitudes $a_q(\varphi)$ and $a_{q+n}(\varphi)$, which predicts that the measured TC corresponds to the dominant OAM contribution within the detected spectral window.

\subsection*{Changes in the topological charge of the detected harmonic radiation}

The measured TC is the azimuthal winding of $E_w$ on a circle of fixed radius $\rho$. Absorbing the radial profiles into two
complex coefficients, the two-mode field can be simplified to: 
\begin{equation}
  E_w(\phi)=c_1 e^{im_1\phi}+c_2 e^{im_2\phi},
  \label{eqn16}
\end{equation}
with $m_1=ql$, $m_2=(q+n)l$ ($m_2>m_1$ for $l>0$), and:
\begin{eqnarray}
  c_1&=&a_q(\varphi) R_{ql}(\rho), \nonumber\\
  c_2&=&a_{q+n}(\varphi) R_{(q+n)l}(\rho) e^{i\Delta\phi_{\rm rel}(\varphi)}. \label{eqn17}
\end{eqnarray}
Using the formal definition to calculate the TC of the detected harmonic radiation $E_w$ over one azimuthal loop, we have:
\begin{equation}
  \ell =\frac{1}{2\pi}\oint_0^{2\pi}\partial_\phi\arg E_w d\phi.
  \label{eqn18}
\end{equation}
Now, we can factorize the helical phase of the $c_1$ component, which leads to: 
\begin{equation}
E_w=e^{i m_1\phi} g(\phi), \label{eqn19} 
\end{equation}
with $g(\phi)\equiv c_1+c_2 e^{i\Delta m \phi}$ and $\Delta m\equiv m_2-m_1=nl$, gives $\ell =m_1+W[g]$, where: 
\begin{equation}
W[g] = \frac{1}{2\pi} \times \oint_0^{2\pi}\partial_\phi\arg{g(\phi)} d\phi. \label{eqn20}
\end{equation}
is the TC of the component $g(\phi)$. Consequently, we can write the measured TC as:
\begin{equation}
      \ell =
  \begin{cases}
    q\times l, & |a_q| |R_{ql}(\rho)|^{p}>|a_{q+n}| |R_{(q+n)l(\rho)}|^{p},\\[2pt]
    (q+n)\times l, & |a_{q+n}| |R_{(q+n)l}(\rho)|^{p}>|a_q| |R_{ql}(\rho)|^{p}.
  \end{cases}
  \label{eqn21}
\end{equation}
The measured TC is determined by the dominant order in the spectrum of the detected window. In the few-cycle regime, the sub-cycle field asymmetry, which is set by the CEP, redistributes the spectral weight between the two overlapping orders. The window amplitudes are therefore genuinely CEP dependent, $|a_q(\varphi)|$ and $|a_{q+n}(\varphi)|$, and their ratio varies over a CEP cycle as $\varphi$ tunes the balance between the two modes.



\section*{Acknowledgments}
The ELI ALPS project (GINOP-2.3.6-15-2015-00001) is supported by the European Union and co-financed by the European Regional Development Fund. C.G. and B.K.D. acknowledge the support of the Extreme Light Infrastructure ERIC (ELI-ERIC) through the program ``ELI Call for Users". W. Gao thanks financial support from National Natural Science Foundation of China (Grant number: 12474309). C.R.G. acknowledge support from SECIHTI, México (CBF-2025-I-1804).


\section*{Author contributions statement}
C.G. and R.S. equally contributed to this work. C.G., D.R., B.K.D., B.K. and R.S. made the experimental work. C.G., B.K.D., B.K., D.R. and E.C. conceptualized the idea. C.G. and W.G. prepared the materials for the experiments. C.G., R.S. and B.K.D. wrote the manuscript. C.G., CRG and W.G. supervised the project. All authors reviewed and corrected the manuscript. 

\section*{Conflict of Interest Statement}
The authors declare no conflict of interest. 

\bibliography{sample}

\clearpage

\section{Supplementary material}
The supplementary material is aimed to support the results presented in the main manuscript and present additional 
theoretical and experimental results. We will start by describing a superposition of fundamental modes followed by the superposition of harmonic vortices. We then will present the high-order harmonic generation (HHG) simulations in solid including the symmetry breaking mechanism. Finally, we will present further experimental measurements to support our main findings.

\section{Superposition of vortex beams}

\begin{figure*}[ht!]
    \centering
    \includegraphics[width=1\columnwidth,trim=0cm 0cm 0cm 0cm, clip]{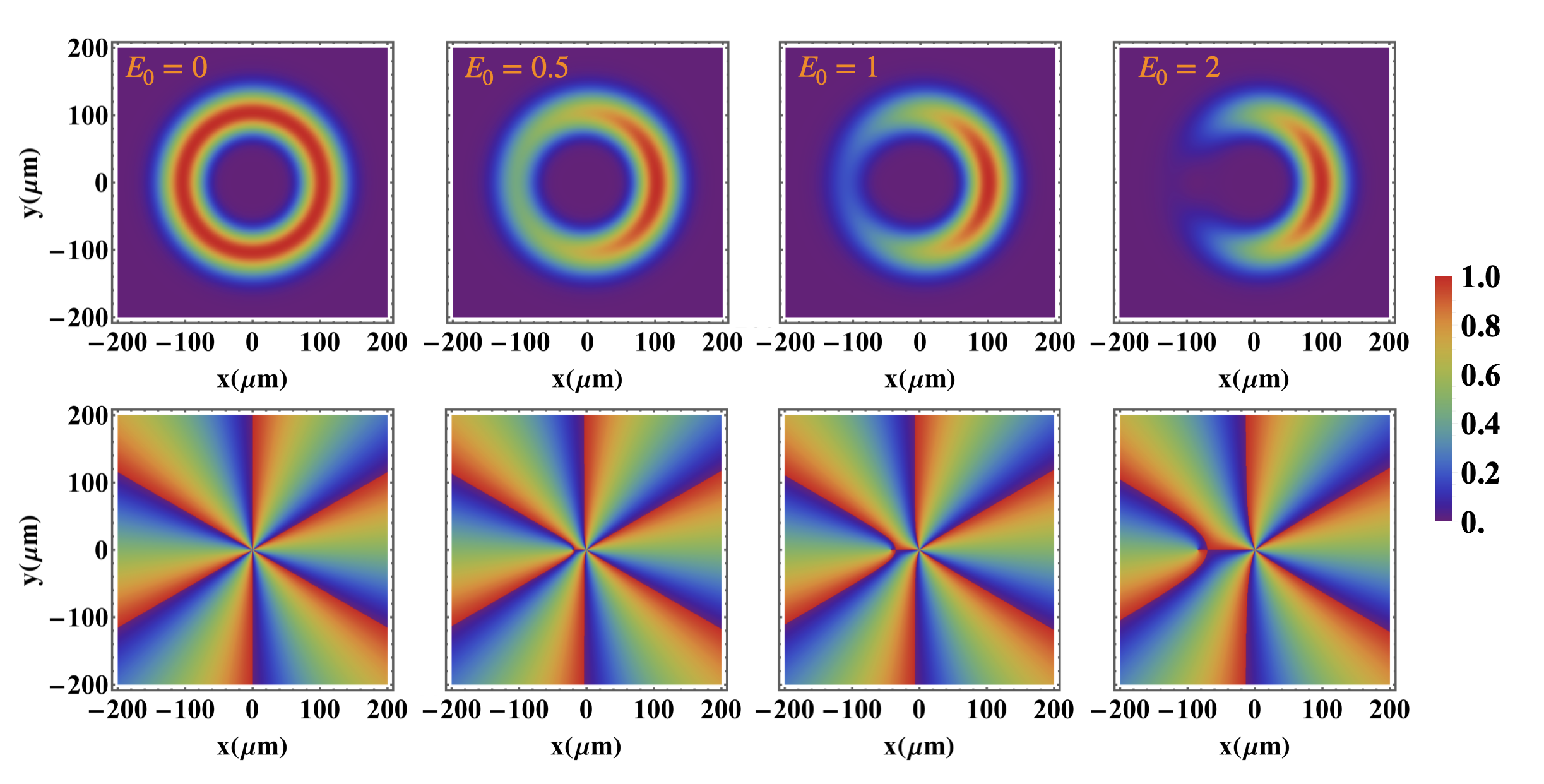}
    \caption{Vortex beam and phase plots. The characteristic ‘‘c" patter can be observed when one of the beams has twice the field amplitude. This pattern coincide with the experimental results. Importantly, the ‘‘c" patter can become more open (smaller intensity maxima), if there are more than one OAM mode interfering.}
    \label{Fig8}
\end{figure*}

The super of two Laguerre-Gauss beams (LGBs) give rise to intensity distributions which do not conserve the rotational symmetry. To simulate the superposition we will write the LGB as 

\begin{equation}
    E_{LG}(r,\phi)=E(r)\times e^{il\phi}
\end{equation}
with, $l$ the topological charge (TC), $\phi=\arctan(y/x)$ and $E(r)$ the spatial distribution of the vortex beam. The superposition of two different modes can be written is
\begin{eqnarray}
    E(r,\phi) &=& E^1_{LG}(r,\phi)+ E^2_{LG}(r,\phi) = E_1(r)\times e^{il_1\phi}+E_2(r)\times e^{il_2\phi}\nonumber \\
    &=& e^{il_1\phi}\big(  E_1(r) + E_2(r)\times e^{i(l_2-l_1)\phi} \big).
\end{eqnarray}
Is the term $\exp{(i(l_2-l_1)\phi)}$, that breaks the rotational symmetry, i.e., $E(r,\phi+\alpha)\ne E(r,\phi)$, since the intensity acquires a term that depends on the transformation angle $\alpha$. Consequently, the superposition of modes naturally results in vortex beams with no rotational symmetry. Importantly, this results depends on the relative amplitude between the modes and their TC difference. We simulated the superposition of two beams, as shown in Fig.~\ref{Fig8}. The results demonstrate the effect of the interference between the two beams. In the calculations we kept the amplitude of one field constant and changing the amplitude of the second one, we arrive to the symmetry broken vortex beams shown in Fig.~\ref{Fig8}. Furthermore, the effect of the broken symmetry is observed in the phase calculations, which shown a dislocation along the $y=0$ axis. 

\begin{figure*}[ht!]
    \centering
    \includegraphics[width=1\columnwidth,trim=0cm 0cm 0cm 0cm, clip]{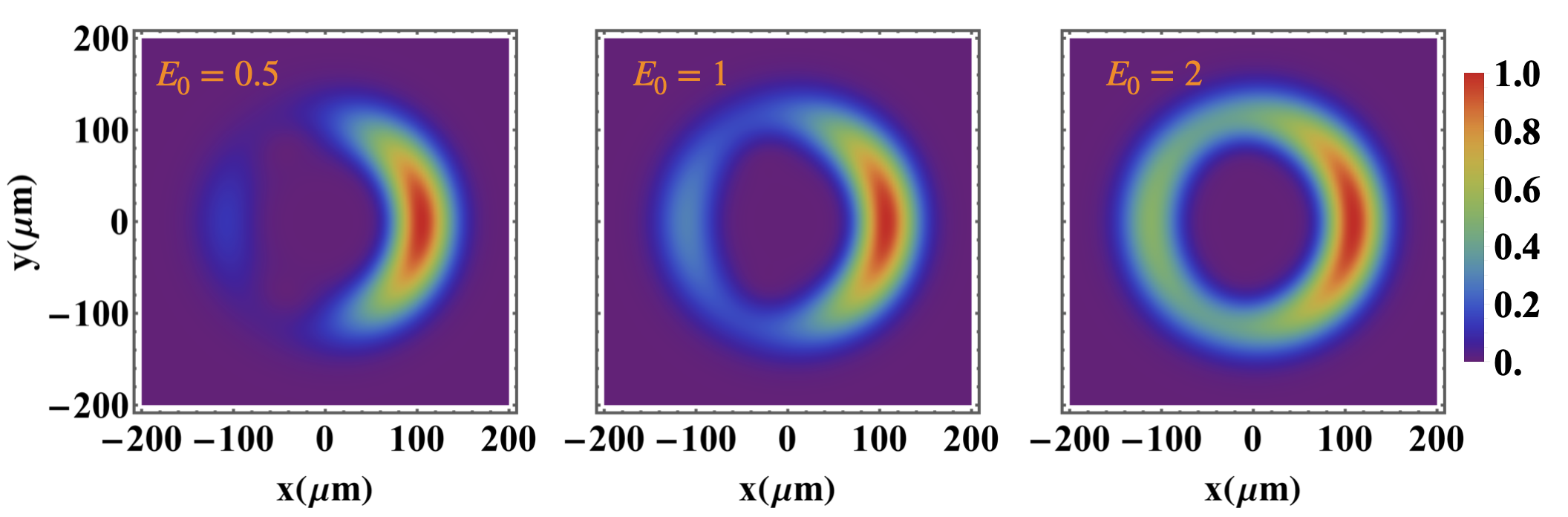}
    \caption{Inteference of three different vortex beams with different TC. }
    \label{Fig9}
\end{figure*}

It is important to note that, the superposition changes the vortex beam intensity profile as well as the phase plot if there are more modes involved. For example the superposition of three modes results in a more open "c" pattern or even in two maxima, depending on the ratio of their amplitude. An example of this situation is shown in Fig.~\ref{Fig9}, where three vortex beams with different TC interfere. 

The results presented here demonstrate that there is a multimode superposition in the measured bandwidth. Most importantly, the results coincide with the experimentally measured harmonic vortex beams presented in the main manuscript. 

To demonstrate that we also generate fractional TCs, we calculated the OAM value of the superposition along the $z$-direction, which is the beam propagation direction. For this, we calculated the expectation value $\langle L_z \rangle$, with the two modes OAM eigenstates $|\psi\rangle = a|l\rangle + b|(l+1)\rangle$. Notice that we assume the TC values to be different by one sine in the experiment we detected the harmonic radiation between two harmonics. 

\begin{equation}
    \frac{\langle \psi| L_z|\psi \rangle}{\langle \psi| \psi\rangle} =\hbar  \frac{|a|^2 l +|b|^2(l+1)}{|a|^2+|b|^2}.
\end{equation}
Here, $a$ and $b$, represent the beam amplitudes. For the case of two orthogonal OAM states, the expectation value for two interfering modes with the same amplitude results in
\begin{equation}
    \frac{\langle \psi| L_z|\psi \rangle}{\langle \psi| \psi\rangle} =\hbar \Bigg( l+\frac{1}{2}\Bigg).
\end{equation}
which results in a fractional value. Most importantly, the interference between the OAM modes is only spatial, since the OAM eigenstates are orthogonal. The OAM expectation value contains no interference for the same reason $\langle l| (l+1)\rangle = 0$. 

The interference of three different beams is more interesting. If the beams have TC values $l$, $l+1$ and $l+2$, and the same field amplitude, the TC of the coherent superposition is an integer value. However, if the beams have unequal amplitudes, the expectation value $\langle L_z\rangle$ is fractional. The result can be geenralize as follows

\begin{equation}
    \frac{\langle \psi| L_z|\psi \rangle}{\langle \psi| \psi\rangle} =\hbar \frac{\sum_l l |A_m|^2}{\sum_l |A_m|^2},
\end{equation}
which results from the orthogonality of the OAM eigenstates. 

Most importantly, the CEP effect is exactly that: it changes the relative amplitude between the superimposed modes resulting in different net TC values. However, the main modes, experience no change in the TC. 

\section{Other vortex beam and topological charge measurements}

\begin{figure*}[ht!]
    \centering
    \includegraphics[width=1\columnwidth,trim=0cm 0cm 0cm 0cm, clip]{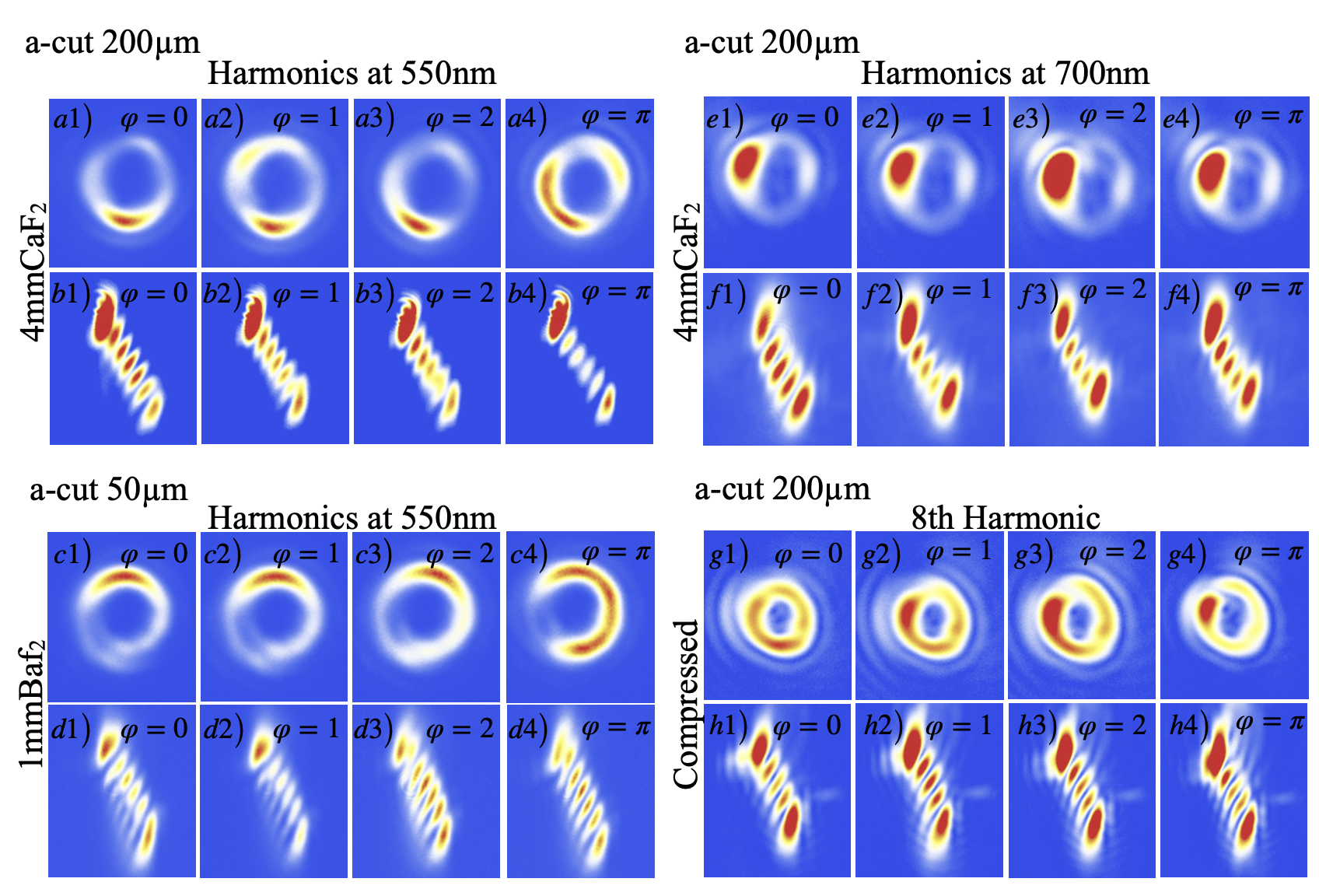}
    \caption{Beam profile and topological charge measurements. In panels (a) and (b), we show the beam and HG profiles for 4 different CEP values. The evolution of the HG profile shows a change in the TC. For thinner aZnO crystal, the CEP effect is confirmed,, as shown in panels (c) and (d). The CEP effect is reduced for the spectral region around 700~nm, which lies between the fourth and fifth harmonic vortices, as shown in panels (e) and (f). Interestingly, for the 4$^\text{th}$ harmonic, we also observe several modes, as shown in panels (g) and (h). The other mode present in this harmonic is more probably the 3$^\text{rd}$ harmonic. The changes in the harmonic vortices are evident in the dislocation present in the vortex beam for $\varphi = \pi$ and the HG profile for $\varphi = \pi$. }
    \label{Fig6}
\end{figure*}

The CEP scans shown in Fig.~2 of the main manuscript, demonstrate the CEP effect on the harmonic spectra. Additionally, we presented in Fig.~3 of the main manuscript, the HG lobes changes between $\varphi = 0$ and $\pi$. This values represent the strongest changes in the vortex beam intensity distribution and the TC measurements. However, some changes can be observed in other CEP phase values. To complement the measurements in the main manuscript, we present in Fig.~\ref{Fig6}, the measured vortex beam and HG lobes (to extract the TC) for different crystals and spectral regions. In panels (a1) to (a4) we show the vortex beam for the region between harmonics 5$^\text{th}$ and 6$^\text{th}$ and generated in a a-cut crystal with a thickness of 200~$\mu$m. For the four different CEP values there are changes not only in the intensity distribution of the beam but also in the vortex beam structure itself. The composition of several modes in the detected radiation creates a fractional vortex beam, which during propagation tries to acquire the closes integer TC. The evolution of the HG lobes for different CEP values is presented in Fig.~\ref{Fig6} (b1) to (b4). For a CEP values $\varphi=0$ the detected radiation exhibit a TC value of $l_q = 6$, while for a $\varphi=\pi$, the TC changes to $l_q = 5$. This is clear from the number of minima between the lobes in the HG pattern. Interestingly, for the short pulse, the effect is not so clear. That is why 

Similar results were found for the 50~$\mu$m thick a-cut crystal, as shown in Fig.~\ref{Fig6} (c1) to (c4) for the vortex beam and in (d1) to (d4) for the HG pattern. 
Interestingly, the vortex beam in (c1) to (c4) exhibits a double structure around dark core, which evolves, as we change the TC to an almost open ring, which clearly demonstrate the superposition of modes. Moreover, between panel (d1) with $\varphi =0$ and panel (d4) with $\varphi = \pi$ there is a change of TC value of 1. However, the effect of the chirp in the driving field is evident here: for the 50~$\mu$m thick crystal it is necessary to stretch the driver with a 1~mm BaF$_2$ window, while for the  200~$\mu$m thick crystal, it is necessary to use a 4~mm CaF$_2$ window. 

Additionally, we measured the CEP effect for the spectral region around 700~nm, which lies between the 5$^\text{th}$ and 3$^\text{rd}$ harmonics. In Fig.~\ref{Fig6} (e1) to (e4) we show the measured vortex beam while in panels (f1) to (f4) we show the measured HG pattern. Changes as a function of the CEP are observed for both measurements, however without a change in the TC. 

We also measured the changes in the exact harmonics. In Fig.~\ref{Fig6} (g1) to (g4), we show the 8$^\text{th}$ harmonic vortex beam, while in panels (h1) to (h4) we show the measured HG patters. Here, it is clear that there are several modes interfering. Even when there is not a net change in the TC value, the multimode nature of the harmonic is evident for $\varphi=\pi$, where we can observed a beam dislocation. The beam evolves from a symmetric vortex beam to a fractional one. This can also be observed in the HG pattern for the same CEP value.

\section{Long pulse measurements}

\begin{figure*}
    \centering
    \includegraphics[width=1\columnwidth,trim= 0cm 0cm 0cm 0cm, clip]{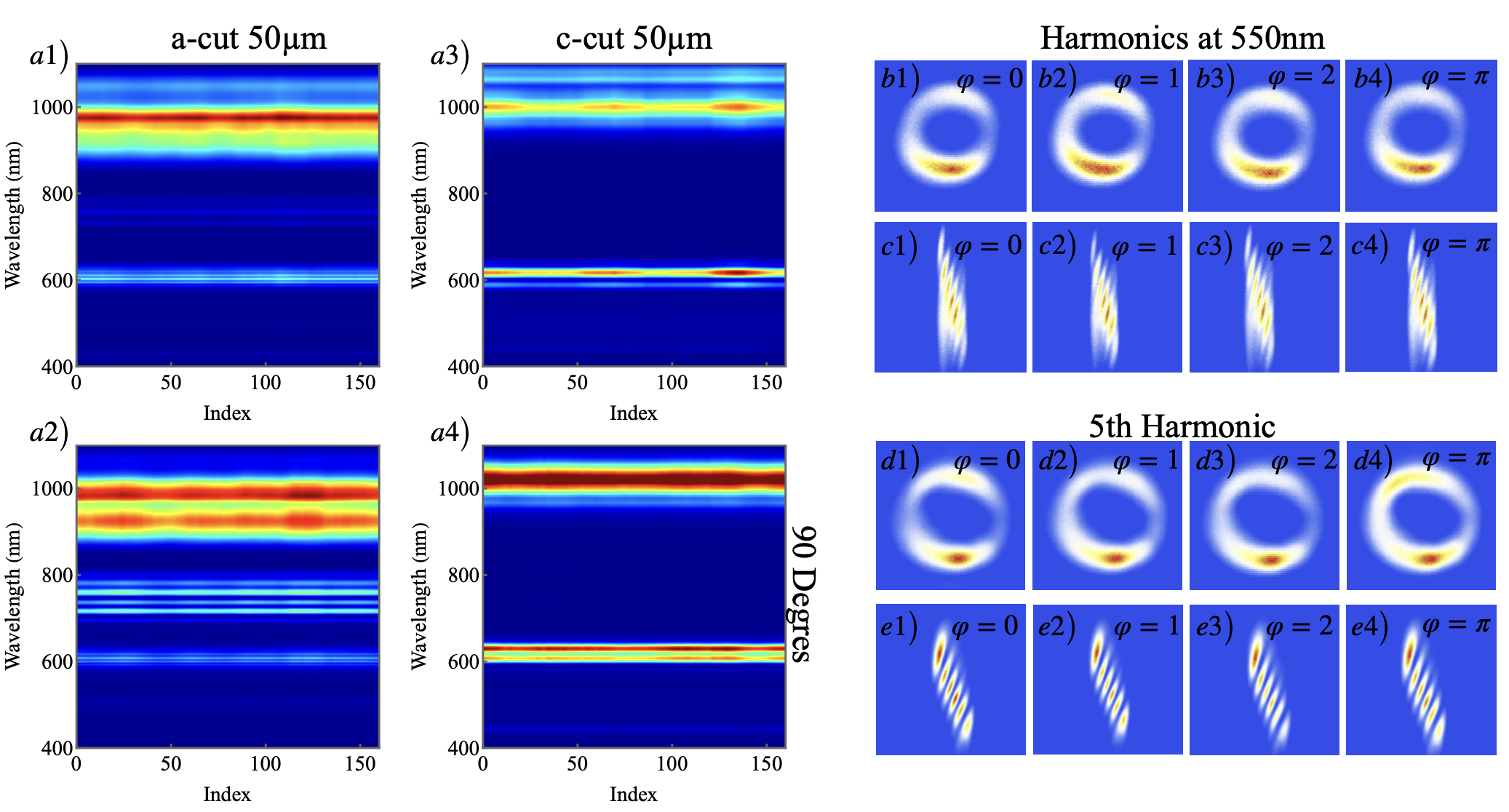}
    \caption{Long pulse measurements. In panels (a) we show the CEP measurements for both aZnO and cZnO for two different angles. The effect of the CEP is not present in the scans. Furthermore, for the spectral region around $550$~nm and the fifth harmonic there is not CEP effect, as shown in panels (b), (c), (d) and (e).}
    \label{Fig7}
\end{figure*}

Further confirmation of the CEP effect on the detected harmonic radiation is found in Fig.~\ref{Fig7}. In panels (a1) to (a3), we present the 4 consecutive CEP scans for the aZnO and cZnO and for 0 and 90$^\text{o}$ rotation angle of the solid sample. Contrary to the cases presented in Fig.~\ref{Fig2}, for long pulses there is no CEP effect for either the spectral region in between the harmonic or for the harmonics itself. In particular, in for the 550~nm region and the 5$^\text{th}$ harmonic, as shown in panels (b), (c), (d), and (e), there is not CEP effect. It is important to note that, because the long pulses, there is almost not harmonic radiation been produced at 550~nm. This implies that the measurements presented in panels (b) and (c) were saturated to obtain clear images. However, for the HG profiles, it was not possible to remove the noise due to the low signal. Also, notice that there are not changes in the HG profile or the noise level, indicating no CEP effect, as in the previous cases.

In conclusion, our measurements and theoretical studies clearly demonstrate the microscopic origin of the CEP effect on the superposition of different OAM modes. Even when the most notorious effect happens in spectral regions where there are not harmonics, the short pulses create a more continuous harmonic spectra. This allow us to select part of the spectra where for long pulses there is not harmonic generation. In this regions, because the mode spectral distance is small, the CEP effect can be observed. 

This experimental observations open the door to investigate the harmonic plateau, where the spectra separation between consecutive harmonics is even shorter. The CEP control of the TC of particular harmonics in that particular region will be possible making TC sensitive measurements feasible.

\section{Semiconductor-Bloch equations simulations}

To demonstrate the CEP effect in the harmonic spectra, we model the electron dynamics in a one-dimensional Brillouin zone (BZ) with crystal momentum $k\in[-\pi/a,\pi/a]$. The electromagnetic field is given by 
\begin{eqnarray}
E(t)&=&E_0\, f(t)\sin(\omega_0 t+\varphi),
\nonumber \\ 
f(t)&=&\sin^2\!\left(\frac{\omega_0 t}{2 n_c}\right),
\end{eqnarray}
for $0\le t\le n_c T_0$ (and $f(t)=0$ otherwise), where $T_0=2\pi/\omega_0$ is the
optical period, $n_c$ the number of cycles, and $\varphi$ the carrier-envelope phase. The valence and conduction band dispersions are modeled by
\begin{align}
\varepsilon_v(k) &= \sum_{n=0}^{5} a_{nx}^{v}\cos(nka_x),\\
\varepsilon_c(k) &= E_g+\sum_{n=0}^{5} a_{nx}^{c}\cos(nka_x).
\end{align}
The expansion coefficients were taken from Ref.~\cite{...}. The conventional SBEs can be written as follows
\begin{eqnarray}\label{sbes}
i\frac{\partial}{\partial t}p_k(t)&=&\left(\varepsilon_k^e+\varepsilon_k^h-i\frac{1}{T_2}\right) p_k(t)\nonumber \\
&&-(1-n_k^e-n_k^h)E(t)d_k-iE(t)\nabla_k p_k(t) \nonumber\\
\frac{\partial}{\partial t}n_k^e(t)&=&-2\mathrm{Im}[ E(t)d_k p_k^{*}]-E(t)\nabla_k n_k^e\nonumber \\
\frac{\partial}{\partial t}n_k^h(t)&=&-2\mathrm{Im}[ E(t)d_k p_k^{*}]-E(t)\nabla_k n_k^h.
\end{eqnarray}
\\
Here, $\varepsilon_k^e=\varepsilon_c(k)$ is the electron (conduction-band) energy and
$\varepsilon_k^h=-\varepsilon_v(k)$ the hole energy, so that the band gap is
$\varepsilon_k=\varepsilon_k^e+\varepsilon_k^h=\varepsilon_c(k)-\varepsilon_v(k)$. The transition dipole coupling is represented by $d_k$. In the length gauge the dynamics are described by the microscopic interband coherence $p_k(t)$ and the electron and hole populations $n^e_k(t)$, $n^h_k(t)$, driven by a spatially uniform field $E(t)$. Furthermore, the terms $-E(t)\nabla_k p_k(t)$, $-E(t)\nabla_k n_k^e$, and $-E(t)\nabla_k n_k^h$ account for intraband acceleration in crystal momentum space. From these quantities we calculate the polarization $P(t)$ and intraband current density $J(t)$
\begin{eqnarray}\label{polarization}
 P(t)&=&\sum_{k}\left[ d_k\;p_k(t)+\mathrm{c.c}\right],\\   
 J(t)&=&-2\sum_{k}\left[ v_k^{e}n_k^e(t)+v_k^{h}n_k^h(t)\right],
\end{eqnarray}
here, the group velocities are $v_k^{e,h} = \nabla_k\varepsilon_k^{e,h}$. Additionally, the interband and intraband contributions to the total current are
\begin{eqnarray}
J_{\rm inter}(t)&=&\frac{dP(t)}{dt},\\
J_{\rm intra}(t)&=&J(t),
\end{eqnarray}

which are added to obtain the total current

\begin{equation}
J_{\rm tot}(t)=J_{\rm inter}(t)+J_{\rm intra}(t),
\end{equation}

from which the HHG spectrum is calculated as

\begin{equation}
I_{\rm HHG}(\omega)=
\left|F[J_{\rm tot}(t)]\right|^2.
\end{equation}

\begin{figure*}[ht!]
    \centering
    \includegraphics[width=1\columnwidth,trim=0cm 0cm 0cm 0cm, clip]{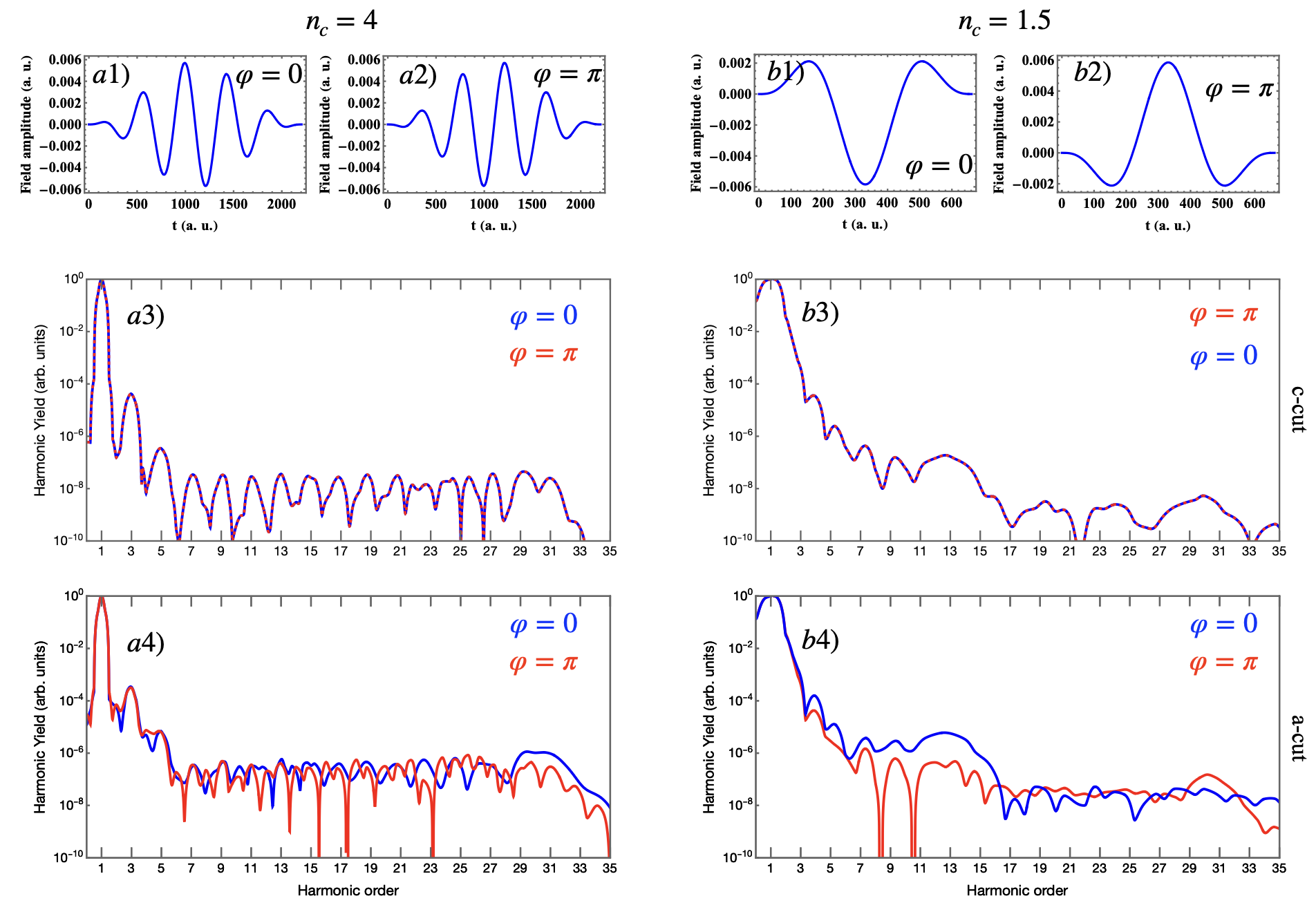}
    \caption{Calculated harmonic spectra for a pump field with 4 and 1.5 cycles. In (a1) and (a2), we present the simulated pump driving field for $\varphi = 0$ and $\varphi = \pi$ and $n_c=4$, respectively. In (a3) and (b4), we show the harmonic spectrum for the same CEP phases and for cZnO and aZnO, respectively. The corresponding results for $n_c=1.5$ are shown in Figs. (b1) to (b4). A clear CEP dependence is observed for the aZnO, interestingly for both long and short pulses. More importantly, between harmonics orders 5$^\text{th}$ and 7$^\text{th}$ there is clear CEP effect and an indication of a superposition of modes. }
    \label{Fig5}
\end{figure*}

Now, we can include the spatial inversion symmetry breaking into the 1D-SBEs model. For this, we will follow the theoretical developments presented in Ref.~\cite{EvenH_dipoleZnO,EvenH_dipoleGraphene}, where it was demonstrated that it is possible to relate the observation of even harmonics to the phase of the transition dipole moment. In practical terms this means the introduction of a complex transition dipole moment \cite{EvenH_dipoleGraphene} in the SBEs to solve the electron dynamics in the solid. To achieve this, we re-write the dipole moment in the form 
\begin{equation}\label{eq:complexd}
d(k) = |d(k)| e^{i\Phi(k)},
\end{equation}
with $\Phi(k)$ the dipole phase. Substituting the complex dipole Eq.~\ref{eq:complexd} into the conventional SBEs is not a matter of redefining a coefficient: the real and imaginary parts of the coherence couple differently once $d$ is complex. 

In a smooth local gauge for a two-band Hamiltonian
\begin{equation}
H(k)=h_0(k)\mathbf{1}+\mathbf{h}(k)\cdot\boldsymbol{\sigma},
\end{equation}
the interband Berry connection can be written as
\begin{equation}
d_{cv}(k)
=
i\langle u_c|\partial_k u_v\rangle
=
\frac{1}{2}
\left(
\sin\theta\,\partial_k\gamma
+
i\,\partial_k\theta
\right),
\label{eq:dcv}
\end{equation}
up to an overall gauge-dependent phase/sign convention. The dipole transition moment can be write as a complex number in the form
\begin{equation}
d_{cv}(k)=d_r(k)+id_i(k),
\end{equation}
its phase is
\begin{equation}
\Phi(k) = \arg \bigl[d_{cv}(k)\bigr]= \operatorname{atan2} \bigl( \partial_k\theta, \sin\theta \partial_k\gamma \bigr).
\label{eq:dipolephase}
\end{equation}

Furthermore, writing all the quantities in polar representation of the new complex phase we have 
\begin{eqnarray}
p_k &=& p_k^r + i p_k^i, \nonumber \\
d_k &=& d_k^r + i d_k^i, \nonumber \\
d_k^r &=& |d_k|\cos\phi, \nonumber \\
d_k^i &=& |d_k|\sin\phi,
\end{eqnarray}
and inserting these into Eqs.~\ref{sbes} yields the real-valued system that can be solved numerically in the same way than Eqs.~\ref{sbes}
\begin{eqnarray}\label{eq:sbes2}
\frac{\partial}{\partial t} p_k^r(t) &=& -\varepsilon_k p_k^i(t) - \frac{p_k^r}{T_2} - E(t)\nabla_k p_k^r(t) \nonumber \\
&-& d_k^i (1 - n_k^e - n_k^h) E(t), \nonumber \\
\frac{\partial}{\partial t} p_k^i(t) &=& -\varepsilon_k p_k^r(t) - \frac{p_k^i}{T_2} - E(t)\nabla_k  p_k^i(t) \nonumber \\
&+& d_k^r (1 - n_k^e - n_k^h) E(t), \nonumber \\
\frac{\partial}{\partial t}n_k^e(t) &=& 2\big(d_k^r p_k^i - d_k^i p_k^r\big)E(t) - E(t)\nabla_k n_k^e, \nonumber\\
\frac{\partial}{\partial t}n_k^h(t) &=& 2\big(d_k^r p_k^i - d_k^i p_k^r\big)E(t) - E(t)\nabla_k n_k^h.
\end{eqnarray}
here, $\varepsilon_k = \varepsilon_k^e+\varepsilon_k^h$. The new complex transition dipole are the new additions relative to the conventional model. Setting $d_i\to0$ (real dipole) collapses
Eqs.~\ref{eq:sbes2} back to the original equations and recovers the
odd-only $c$-cut limit, which is a convenient built-in check. Importantly, for the calculations, we used a phenomenological dipole phase of the form $\Phi(k)=\phi_0 \sin(k a_x)$. The interband polarization Eq.~\ref{polarization} was evaluated with the complex dipole as follows
\begin{equation}
P(t) = 2\mathrm{Re}\Bigg(\sum_k (d_r + i d_i)(p_r + i p_i)\Bigg).
\end{equation}
An example of the numerical solutions of Eq.~\ref{eq:sbes2} is presented in Fig.~\ref{Fig5}. In (a1) and (a2), we show the driving field for a number of cycles $n_c=4$ and a CEP phase of $\varphi=0$ and $\pi$, respectively. In (b1) and (b2) we show the driving field compose of $n_c=1.5$ and for the same CEP phase, $\varphi=0$ and $\pi$, respectively. The simulated HHG spectra for the cZnO, labeled as "c-cut", is presented in (a3) for the long and (b3)  and short driving field. In (a4) and (b4) we show the results for aZnO, labeled as "a-cut" for the long and short number of cycles, respectively. As the results demonstrate, when we introduce the complex transition dipole moment. the CEP sensitivity appears, while for the real transition dipole moment, the results demonstrate no sensitivity to the CEP. The CEP effect in the harmonic spectra supports our hypothesis explaining the microscopic origin of the OAM control: Changing the CEP, changes the emission window of the harmonics. Microscopically this means that, we changes the relative contribution and phase of the emitted harmonics. As a consequence the interference between near by orders changes and for the detected wavelength, the weight of the superimpose harmonics changes. The strongest mode then imprints the TC of the detected harmonic radiation. 

Even when the aZnO for long pulses also exhibits a CEP sensitivity, there is not a drastic change in the amplitudes of the harmonic orders 5$^\text{th}$ to 7$^\text{th}$. However, for the short pulse ($n_c=1.5$) the changes in their amplitude are evident, further supporting our explanation of the observed CEP effect.

\end{document}